\documentclass{ametsocV6.1}

\usepackage{subcaption}
\graphicspath{{fig/}{../fig/}}

\nolinenumbers

\title{A Bayesian Approach to Atmospheric Circulation Regime Assignment}

\authors{\thanks{This Work has been accepted to Journal of Climate. The AMS does not guarantee that the copy provided here is an accurate copy of the Version of Record (VoR)} \thanks{}
Swinda K.J. Falkena,\aff{a}\correspondingauthor{Swinda K.J. Falkena, s.k.j.falkena@uu.nl} 
Jana de Wiljes,\aff{a,b} 
Antje Weisheimer,\aff{c,d} 
 and Theodore G. Shepherd,\aff{e} 
}

\affiliation{\aff{a}{Department of Mathematics and Statistics, University of Reading, Reading, UK} \thanks{Swinda K.J. Falkena's current affiliation: IMAU, Utrecht University, Utrecht, The Netherlands.} \\
\aff{b}{Institute for Mathematics, University of Potsdam, Potsdam, Germany}\\
\aff{c}{European Centre for Medium-Range Weather Forecasts (ECMWF), Reading, UK}\\
\aff{d}{National Centre for Atmospheric Science (NCAS), University of Oxford, Department of Physics, Atmospheric, Oceanic and Planetary Physics (AOPP), Oxford, UK}
\aff{e}{Department of Meteorology, University of Reading, Reading, UK}
}

\abstract{The standard approach when studying atmospheric circulation regimes and their dynamics is to use a hard regime assignment, where each atmospheric state is assigned to the regime it is closest to in distance. However, this may not always be the most appropriate approach as the regime assignment may be affected by small deviations in the distance to the regimes due to noise. To mitigate this we develop a sequential probabilistic regime assignment using Bayes Theorem, which can be applied to previously defined regimes and implemented in real time as new data become available. Bayes Theorem tells us that the probability of being in a regime given the data can be determined by combining climatological likelihood with prior information. The regime probabilities at time $t$ can be used to inform the prior probabilities at time $t+1$, which are then used to sequentially update the regime probabilities. We apply this approach to both reanalysis data and a seasonal hindcast ensemble incorporating knowledge of the transition probabilities between regimes. Furthermore, making use of the signal present within the ensemble to better inform the prior probabilities allows for identifying more pronounced interannual variability. The signal within the interannual variability of wintertime North Atlantic circulation regimes is assessed using both a categorical and regression approach, with the strongest signals found during very strong El Ni\~no years.} 

\begin{document}

%% Necessary!
\maketitle

%%%%%%%%%%%%%%%%%%%%%%%%%%%%%%%%%%%%%%%%%%%%%%%%%%%%%%%%%%%%%%%%%%%%%
% SIGNIFICANCE STATEMENT/CAPSULE SUMMARY
%%%%%%%%%%%%%%%%%%%%%%%%%%%%%%%%%%%%%%%%%%%%%%%%%%%%%%%%%%%%%%%%%%%%%
%
% If you are including an optional significance statement for a journal article or a required capsule summary for BAMS 
% (see www.ametsoc.org/ams/index.cfm/publications/authors/journal-and-bams-authors/formatting-and-manuscript-components for details), 
% please apply the necessary command as shown below:
%
% Significance Statement (all journals except BAMS)
%
\statement
	 Atmospheric circulation regimes are recurrent and persistent patterns that characterize the atmospheric circulation on timescales of one to three weeks. They are relevant for predictability on these timescales as mediators of weather. In this study we propose a novel approach to assigning atmospheric states to six pre-defined wintertime circulation regimes over the North Atlantic and Europe, which can be applied in real time. This approach introduces a probabilistic, instead of deterministic, regime assignment and uses prior knowledge on the regime dynamics. It allows to better identify the regime persistence and indicates when a state does not clearly belong to one regime. Making use of an ensemble of model simulations, we can identify more pronounced interannual variability by using the full ensemble to inform prior knowledge on the regimes.
%
%% Capsule (BAMS only)
%%
%\capsule
%       Enter BAMS capsule here, no more than 30 words. See \url{www.ametsoc.org/index.cfm/ams/publications/author-information/formatting-and-manuscript-components/#capsule} for details.
% 
%% * * If using twocol mode, you will need to use the commands "twocolsig" and "twocolcapsule" in place of "sig" and "capsule"
%%      to ensure that the text box correctly spans across both columns.

%%%%%%%%%%%%%%%%%%%%%%%%%%%%%%%%%%%%%%%%%%%%%%%%%%%%%%%%%%%%%%%%%%%%%
% MAIN BODY OF PAPER
%%%%%%%%%%%%%%%%%%%%%%%%%%%%%%%%%%%%%%%%%%%%%%%%%%%%%%%%%%%%%%%%%%%%%
%

\section{Introduction}
\label{sec:intro}

A thorough understanding of extra-tropical circulation variability on sub-seasonal timescales is important for improving predictability on these timescales. Improvement of this predictability is of great societal relevance for sectors such as renewable energy. Atmospheric circulation, or weather, regimes can describe this variability by dividing the circulation into a small number of states or patterns \citep{Hannachi2017}. These regimes are recurrent patterns that represent the low-frequency variability in the atmospheric circulation. They have been studied for a long time, starting with papers focusing on their identification \citep[e.g][]{Mo1988,Molteni1990,Vautard1990,Michelangeli1995}, with later research discussing their links with other processes and surface impacts \citep[e.g.][]{Straus2004,Cassou2005,Charlton-Perez2018,VanderWiel2019}.

The most commonly used technique for identifying circulation regimes is $k$-means clustering \citep[e.g.][]{Michelangeli1995,Straus2007,Matsueda2018}. This method separates the phase space into $k$ clusters, where the data within each cluster are similar, but dissimilar between the different clusters. The number of clusters $k$ has to be set a priori, for which several approaches such as a classifiability index \citep{Michelangeli1995} or information criteria \citep{OKane2013} are used. One of the drawbacks of this clustering approach is that it yields a hard, deterministic, assignment of the data to each of the regimes. This means that it is difficult to quantify the uncertainty of the regime assignment, as data close to the regime centre is treated the same as data that is only just (by distance) assigned to that regime.

\begin{figure}
    \centering
    \includegraphics[width=1.\textwidth]{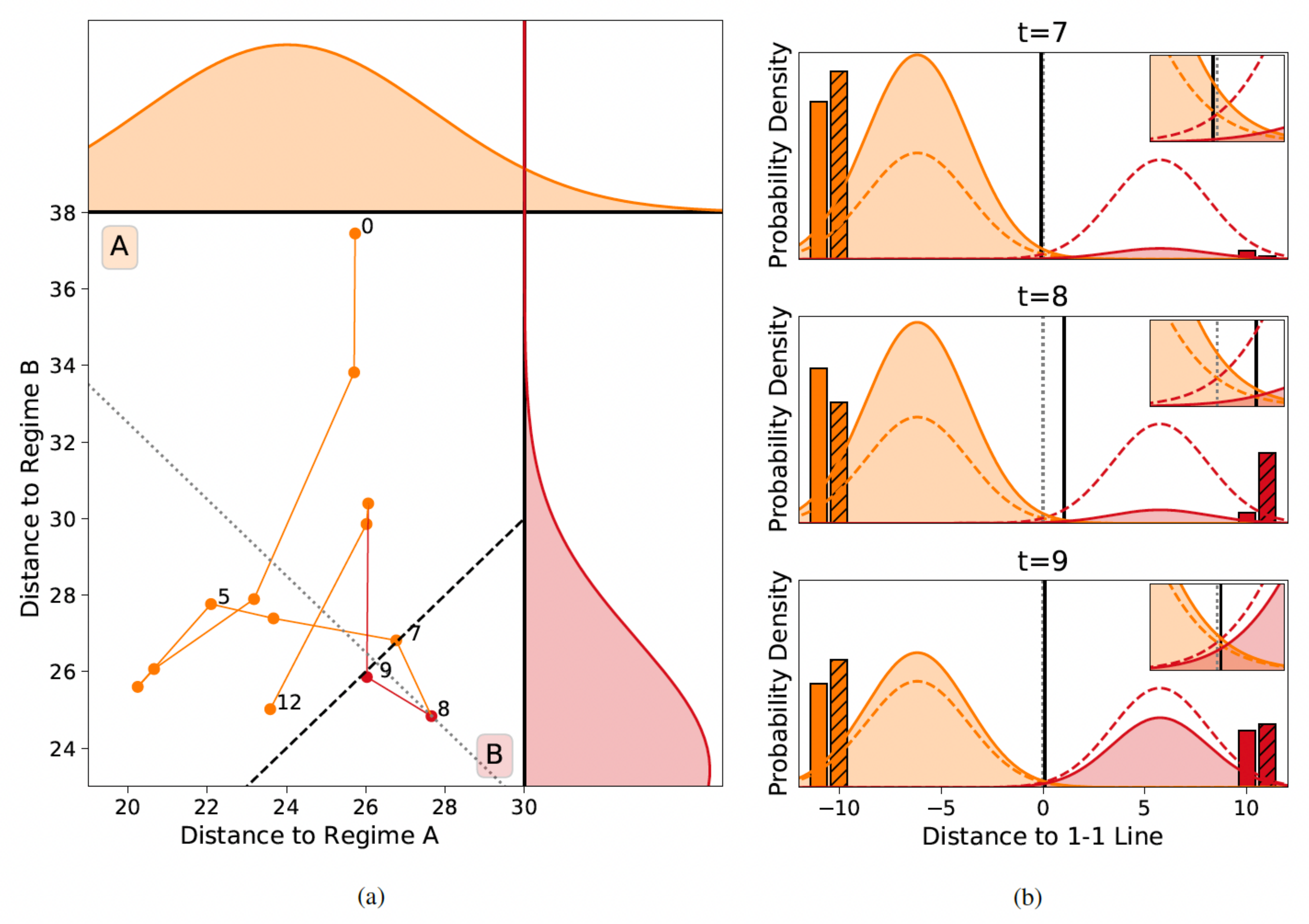}
    \caption{A conceptual example of the difficulty $k$-means clustering has when noise affects the data, showing what a probabilistic approach can bring. (a) An example trajectory of the data as a function of the distances to two regimes A (orange) and B (red). The 1-1 line is shown black dashed, meaning the region above is closer to regime A and the region below to regime B. Numbers indicate the day corresponding to that point in the trajectory. The likelihood functions shown along the top and right give the climatological probability of those distances given hard assignment to regime A (orange, top) or B (red, right). The dotted grey line indicates a slice through the probability space along which the pdfs in panel (b) are considered. (b) A slice of the likelihood functions, weighted by the prior probabilities following Bayes Theorem, for each of the regimes (solid lines, A: orange, B: red) along the grey dotted line in (a), perpendicular to the 1-1 line, for the 7th, 8th, and 9th day. The location of the data on each day is indicated by the vertical black lines, and the bars at the edge of the plots show the prior (left) and posterior (right, hatched) probabilities for each of the regimes (A: orange, left edge, B: red, right edge). The climatological likelihood functions are shown dashed in all panels and the vertical grey dotted line indicates the location of the 1-1 line. The insets in each panel show an enlargement of the region around the 1-1 line.}
    \label{fig:conceptual}
\end{figure}

The hard regime assignment of $k$-means clustering means that the result is susceptible to noise. Consider Figure \ref{fig:conceptual}(a) which shows the distance of the data to two regimes in time for a real case (discussed later in detail), over a period of 12 days. Initially, the data clearly is categorised to belong to regime A, being significantly closer in distance to regime A than to regime B. However, from day 7 to 9 the data makes a brief excursion into a part of the phase diagram that is closer to regime B, after which it moves back to being closest to regime A. The question is whether this is a real signal or simply the effect of noise. Since the regime dynamics is quite persistent in time it is likely to be the latter, but this possibility is not picked up by the hard assignment of a standard $k$-means clustering approach. Often a low-pass filter is applied to remove this high-frequency variability \citep[e.g.][]{Straus2007,Grams2017}, but in \citet{Falkena2020} it was shown that low-pass filtering can lead to a bias in the observed regime frequencies. 

Another solution is to use a regularised clustering algorithm which constrains, or bounds, the number of transitions between the regimes so that it is in line with the natural metastability of the underlying dynamics. 
Such an approach, first introduced in the context of clustering methods by \citet{Horenko2010a}, has for example been applied to discrete jump processes \citep{Horenko2011} with applications in computational sociology \citep{Horenko2011a} and for efficient classification in the context of sparse data settings \citep{Vecchi2022}. In the context of atmospheric dynamics, time regularisation has been used to study the Southern Hemispheric circulation \citep{OKane2013}, the dynamics of the North Atlantic Oscillation \citep{Quinn2021}, and how to identify persistent circulation regimes \citep{Falkena2020}.
A regularised clustering method allows to better identify the signal within the noise, but does require selecting a constraint parameter. This introduces a parameter selection, where e.g. an information criterion is used to decide on a suitable constraint value.

An alternative approach is to make the regime assignment probabilistic rather than deterministic, allowing for a more nuanced and informative regime assignment in the presence of noise. Methods such as mixture modelling provide such a probabilistic regime assignment \citep[e.g.][]{Hannachi2001,Smyth1999,baldo2023}, but are not widely used. 
Hidden Markov Models (HMMs) extend the mixture modelling approach by also taking into account the dynamics of the system and not just the statistics \citep{Majda2006,Franzke2008}, but are hard to fit for relatively short timeseries when the data is high dimensional.
Another approach is to approximate the regime model using local Markov distance functionals with corresponding time dependent probabilities \citep{Horenko2011}.
In studies that look into forecasting of regimes on sub-seasonal timescales, the probability of being in a regime is often considered by looking at the empirical distribution of the (hard) regime assignment across an ensemble \citep{Vigaud2018,Cortesi2021,Bueler2021,Falkena2022}. 
Such an approach is already used in an operational setting by e.g. ECMWF \citep{Ferranti2015}.
A limitation of this method is that it requires availability of ensemble data, where typically the ensemble size is small, and verification is done against a hard regime assignment from reanalysis. 

A probabilistic regime assignment that does not require this availability of ensemble data would help in better assessing the skill in predicting regimes, as it could be applied to reanalysis data which is also subject to noise. Such a regime assignment would allow to identify the instances in which the observations cannot be clearly assigned to one regime or in which a wrong hard assignment is potentially due to noise. This approach allows for a fairer verification of the model by taking some degree of observational uncertainty into account.
Here it is desirable for the approach to be sequential, which allows for the regime assignment to be done in real time making it suitable for operational applications. Most probabilistic regime assignment methods,
such as mixture models or HMMs,
require the availability of the full dataset when computing the regime probabilities, which would mean one has to rerun the clustering algorithm whenever a new datapoint is added. A method that, after training on an initial dataset, can easily be applied to data as it becomes available is more suitable for an operational setting. Such a method can also be applied to predefined regimes, to provide traceability with previous work.

The standard hard regime assignment can be considered as a random process that takes a value in the set of possible regimes at each time. The associated probability can be computed on the basis of metastability frequencies computed from previous or currently available batch data. The aim is to determine the corresponding conditional probability of being in a regime given the data, i.e. $P(\text{Regime}|\text{Data})$. Following Bayes Theorem this is given by
\begin{equation}
    \label{eq:bayes}
    P(\text{Regime}|\text{Data}) = \frac{P(\text{Data}|\text{Regime})P(\text{Regime})}{P(\text{Data})},
\end{equation}
combining prior knowledge of the probability of being in a regime $P(\text{Regime})$ with an observed likelihood given a regime $P(\text{Data}|\text{Regime})$. 
The latter can sometimes be computed from the climatological data.
In Figure \ref{fig:conceptual}(a) the observed (climatological) likelihood functions for both regimes are shown next to the trajectory. The working of Bayes Theorem for such a trajectory is shown in Figure \ref{fig:conceptual}(b), which shows how the inclusion of prior information $P(\text{Regime})$ following Bayes Theorem \eqref{eq:bayes} affects the posterior $P(\text{Regime}|\text{Data})$ for the trajectory at days 7, 8 and 9, following a section along the dotted line in Figure \ref{fig:conceptual}(a). The climatological likelihood functions of the two regimes A and B, indicated by the dashed lines, are weighted (solid lines) using the prior regime probabilities, shown by the non-hatched bars at the edge of the panels. The posterior probabilities are then computed as the values of the weighted likelihood functions at the datapoint (vertical black line). The obtained Bayesian probabilities are indicated by the hatched bars and used to inform the prior probabilities for the next timestep, using climatological information about transition probabilities.

At day 7 the prior information indicates a very high probability of being in regime A as all previous days belonged clearly to that regime. This increases the probability of $t=7$ belonging to regime A and decreases that of belonging to regime B with respect to the climatological likelihood, which would otherwise be evenly balanced between the two regimes. Thus, there is a high probability that the data at day 7 belongs to regime A. Given the known persistence of regimes, the prior information for day 8 again then indicates a high probability of being in this regime, albeit slightly smaller than at $t=7$, which weights the likelihood functions accordingly. Although the data is closer to regime B, the prior information means that there is an approximately equal probability of being in either of the two regimes. The prior for $t=9$ thus does not weight the likelihood functions as much as for $t=7$ and $8$, and thus the data at day 9 being equally close to both regimes means that again the probability of being in either of the regimes is close to a half. This discussion shows how the inclusion of prior information can be used to compute the probability of a regime given the data, and thereby soften the effects of noise, following the fundamental principles of probability as encoded in Bayes Theorem \eqref{eq:bayes}.
As noted above, the approach as discussed here is sequential and can be applied to individual realisations, making it suitable for operational applications. An initial training dataset can be used to obtain the climatological likelihood functions, after which the regime assignment can be applied to data as it becomes available.
The latter regime assignment step is similar to finding the most probable sequence once a HMM is known \citep{viterbi1967,rabiner1989}.

Other aspects than persistence can affect the prior regime likelihood as well. It is likely that non-stationary external factors, such as the El Ni\~no Southern Oscillation (ENSO) or Sudden Stratospheric Warmings (SSWs), have an influence on the prior regime probabilities \citep[e.g.][]{Toniazzo2006,Ayarzaguena2018,Domeisen2020}. The Bayesian approach allows to incorporate such information, either by looking at e.g. an ENSO index or by making use of the availability of ensemble data. In a previous study a regularised clustering method helped to identify a more pronounced interannual regime signal by making use of the information available in an ensemble \citep{Falkena2022}. Similarly, having a more informative prior for Bayes Theorem \eqref{eq:bayes}, incorporating information from external processes, can help in identifying a stronger non-stationary regime signal.
The Bayesian approach discussed here is not the only method in which information on external forcing can be incorporated in the regime assignment \citep[e.g.][]{Franzke2015}, but it is (to our knowledge) the first that allows to do this in a sequential manner.

In this paper we formalise the intuition of Figure \ref{fig:conceptual} and study how to use Bayes Theorem to obtain a probabilistic regime assignment based on predefined regimes
for the wintertime Euro-Atlantic sector. 
The use of predefined regimes respects the scientific value that has already been established for those regimes, e.g. in the relationship with particular climate impacts.
In Section \ref{sec:dataclustering} we discuss the data that are used and the use of standard $k$-means clustering to obtain the circulation regimes that we consider for this study. The two sections that follow explain the way in which Bayes Theorem can be used for the regime assignment,
where an important aim of our work is to link our method to existing work on clustering of circulation regimes.
We start with the most intuitive sequential form (as discussed above) in Section \ref{sec:sequentialbayes} and in Section \ref{sec:ensemblebayes} we consider how the use of ensemble data, which picks up some external forcing signals, can help to update the prior regime probabilities to study interannual regime variability, which is discussed in Section \ref{sec:interannual}. A discussion and conclusion are given in Section \ref{sec:discussion}.

\section{Data and Clustering}
\label{sec:dataclustering}

For the identification of the circulation regimes the 500 hPa geopotential height fields (Z500) from two datasets are used: the ECMWF SEAS5 hindcast ensemble dataset \citep{Johnson2019} and the ERA-Interim reanalysis dataset \citep{Dee2011}. For both datasets, daily (00:00 UTC) gridpoint ($2.5^\circ \times 2.5^\circ$ resolution) Z500 data over the Euro-Atlantic sector (20$^\circ$ to 80$^\circ$N, 90$^\circ$W to 30$^\circ$E) are considered for all winters (DJFM) for which the SEAS5 ensemble data are available (1981-2016). The regimes are computed using gridpoint anomaly data, where the anomalies are computed with respect to the average DJFM climatology  (see \citet{Falkena2020} for the rationale for this choice). 
% The ERA-Interim and SEAS5 anomalies are computed with respect to their own respective climatologies.
Here the climatologies of ERA-Interim and SEAS5 are used as a reference for the computation of their respective anomalies.
The SEAS5 hindcast ensemble has 51 members and is initialised each year on November 1st, which means that by considering data only from December onwards the effect of the atmospheric initial conditions has been effectively lost. This allows us to treat each ensemble member as an alternative, physically plausible yet not observed realisation of the atmosphere \citep{Thompson2017}, subject to the non-stationary influences for that year (notably ENSO). 

\begin{figure}[h]
    \centering
    \includegraphics[width=.8\textwidth]{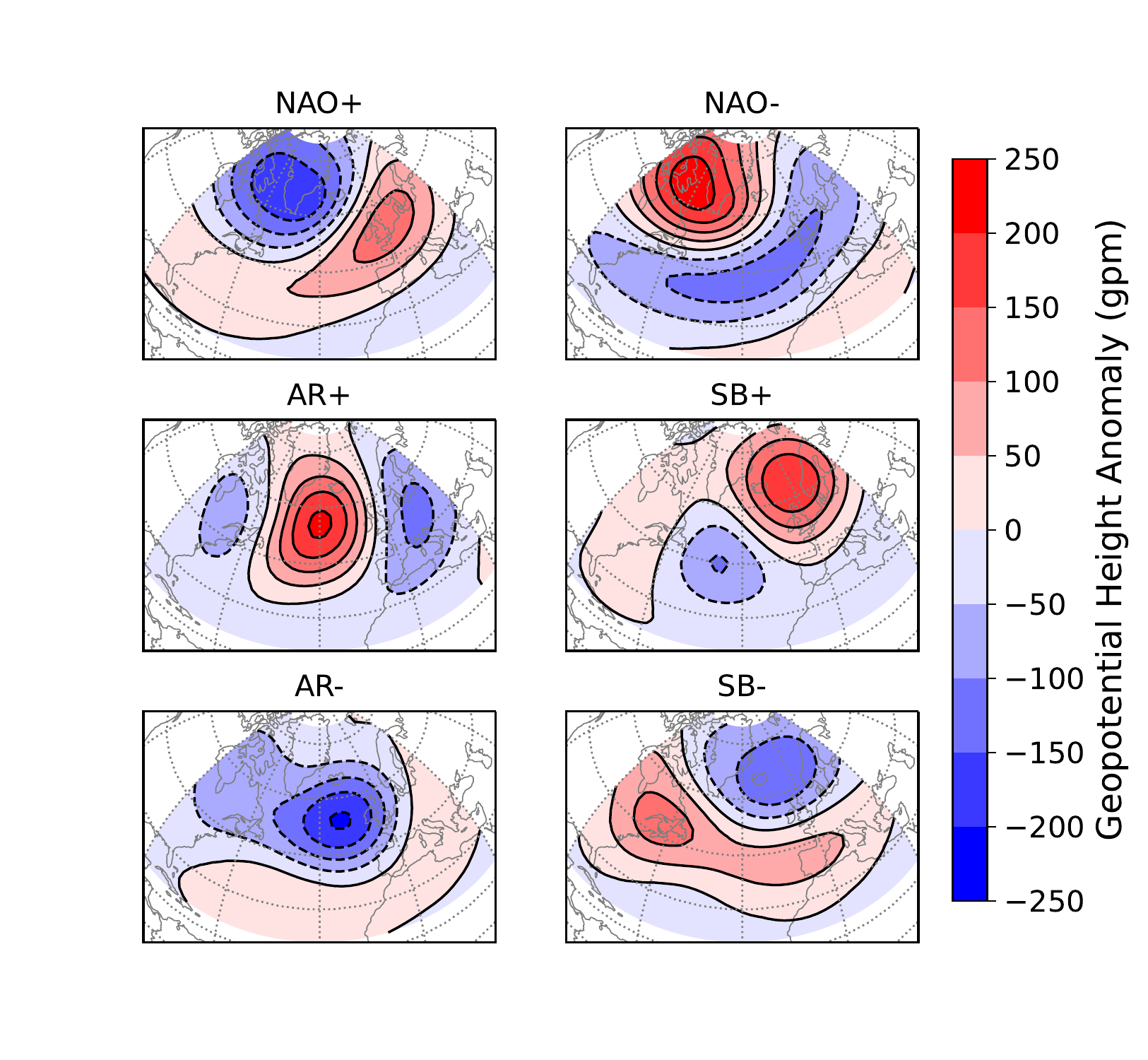}
    \caption{The six circulation regimes obtained for the SEAS5 ensemble using $k$-means clustering. From top-left to bottom-right: 1. NAO+, 2. NAO-, 3. Atlantic Ridge (AR+), 4. Scandinavian Blocking (SB+), 5. AR-, 6. SB-.}
    \label{fig:regimes_ens}
\end{figure}

A standard $k$-means clustering algorithm \citep{Jain2010}, with a Euclidian distance to compute the distance between the data and regimes, is used to identify six circulation regimes over the Euro-Atlantic sector for both ERA-Interim and the SEAS5 hindcast ensemble. 
In $k$-means clustering the data are sorted in $k$ clusters that are close together within one cluster, but far from data in the other clusters based on some distance measure. These clusters are represented by their mean, which corresponds to the circulation regimes, where the number of clusters $k$ has to be set a priori.
Six was identified as a suitable number of regimes for such unfiltered data in a previous study \citep{Falkena2020}. The regimes for the SEAS5 hindcast ensemble are shown in Figure \ref{fig:regimes_ens} and are the two phases of the North Atlantic Oscillation (NAO), the Atlantic Ridge (AR), Scandinavian Blocking (SB) and both their counterparts. Note that these regimes are slightly different in their patterns from those of ERA-Interim (see \citet{Falkena2022} for details on this), thereby providing an inherent bias correction between the model and reanalysis. These hard regime assignments are used to compute the likelihood functions that are used in the Bayesian approach, for which a detailed discussion is given in Section \ref{sec:sequentialbayes}\ref{sec:sequentialbayes_method}. In addition we consider the (hard) regime assignments obtained using the time-regularised clustering algorithm from \citet{Falkena2020}. This allows for a comparison of different approaches to identify the persistent regime signal.

\section{Sequential Bayesian Regime Assignment}
\label{sec:sequentialbayes}

In this section the Bayesian approach to regime assignment is discussed, which can be applied to ERA-Interim data as well as single ensemble realisations. We start with the details of the method itself in Section \ref{sec:sequentialbayes_method}, followed by a comparison with the results of both a standard and time-regularised $k$-means clustering method in Section \ref{sec:sequentialbayes_evaluation}.

\subsection{Method}
\label{sec:sequentialbayes_method}

The starting point for our sequential Bayesian regime assignment is the six regimes obtained using $k$-means clustering discussed in Section \ref{sec:dataclustering} and shown in Figure \ref{fig:regimes_ens}. The likelihood functions in Bayes Theorem \eqref{eq:bayes} are computed based on the distance to these regimes, and remain fixed throughout the sequential Bayesian regime assignment. The discussion of the method as phrased below is general, and can be applied to all types of regime dynamics as long as the regimes themselves and the likelihood functions are specified a priori.

Let $r$ be a discrete random variable indicating a regime, i.e. taking values in $\{1,...,k\}$ for $k$ regimes, and let $\mathbf{d}\in\mathbb{R}^k$ be a vector containing the distances to each of the regimes (here the Euclidian distance is used which is also the standard cost function in the $k$-means setting). Specifically, $\mathbf{d}$ are the data we consider in our Bayesian approach. 
The use of the regime distance as data is not the only option. When one considers only a limited number of principal components (PCs) for the regime representation the PC values can be directly used. However, for the spatial fields considered here (see \citet{Falkena2020} for the arguments in favor of using gridpoint data) this is unfeasible as the high dimensionality means the phase space is sparsely sampled leading to large uncertainty in the resulting distributions.
Therefore, a means of dimension reduction is required for which we consider the distances to the different regimes since this is the metric used in most clustering approaches.
At a given time we are interested in the probability to be in a regime $r$ given the data, i.e. $ P(r|\mathbf{d})$. Bayes Theorem tells us that
\begin{equation}
    \label{eq:bayesth}
    P(r|\mathbf{d}) = \frac{P(\mathbf{d}|r)P(r)}{P(\mathbf{d})}.
\end{equation}
Here, $P(r)$ is the prior probability of regime $r$ and $P(\mathbf{d})$ is the probability of the data. Since we only consider a discrete number of regimes which are mutually exclusive and exhaustive, the latter can be computed by
\begin{equation}
    \label{eq:probdata}
    P(\mathbf{d}) = \sum_{r=1}^k P(\mathbf{d}|r)P(r),
\end{equation}
making it a normalisation factor.

\begin{figure}[h]
    \centering
    \includegraphics[width=1.\textwidth]{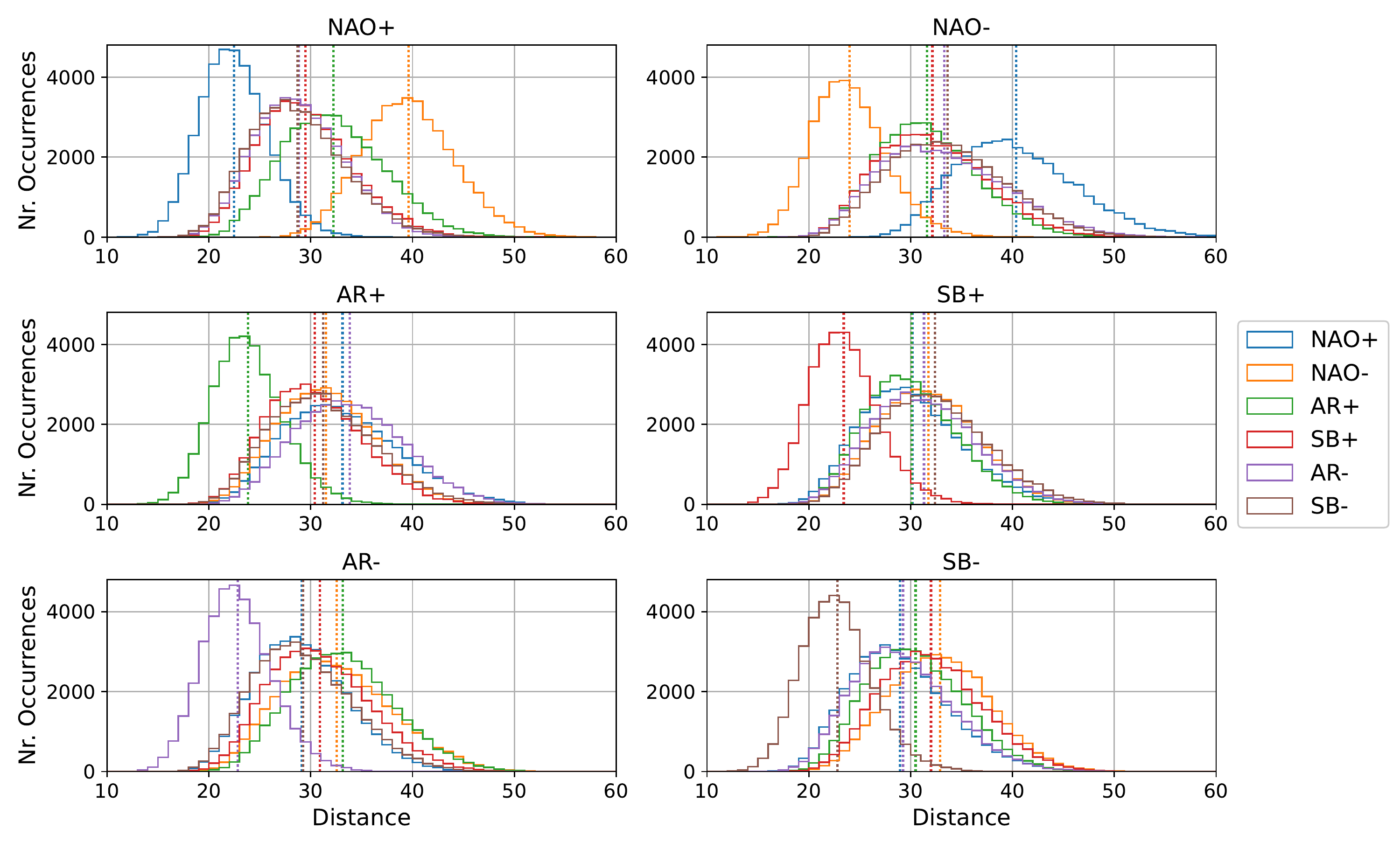}
    \caption{The distributions of the distances (normalized, gpm/\#gridpoints) to each of the regimes (color) conditional on the SEAS5 hindcast data being assigned to the regime given in the title, based on a hard assignment. The means of each distribution are indicated by the vertical dotted lines.}
    \label{fig:distancedistr}
\end{figure}

Lastly, $P(\mathbf{d}|r)$ is the likelihood of the data given a regime $r$. The likelihood of the data can be determined from the distance to each of the regimes by considering how the data fall within the conditional distance distributions, i.e. the distributions conditioned on data belonging to one of the regimes. For each datapoint in either the SEAS5 or ERA-interim timeseries we have this distance to each of the $k$ regimes, which has been computed in the $k$-means clustering procedure to determine the hard regime assignment (Section \ref{sec:dataclustering}). This gives the distributions of the distances to each of the regimes conditional on regime $r$, which for SEAS5 are shown in Figure \ref{fig:distancedistr}. 

There are a few things to note concerning these distributions. Firstly, the distance to the regime the data is assigned to is smallest, but can still be larger than the distance to other regimes for a different datapoint belonging to that regime. Secondly, for data assigned to AR+, SB+, AR$-$ and SB$-$ the distances to the other regimes are roughly equally distributed with the means being relatively close to each other. However, for data assigned to either NAO+ or NAO$-$ the distance to the other phase is larger than that to the other four regimes. Thus these two regimes are further away from each other than the rest of the regimes, and information on the proximity to one regime is providing information on the proximity to the other.

Also, we see that these distributions are approximately normal, justifying us to approximate the corresponding $k$-dimensional conditional probability density functions (pdf) by a multivariate normal. The likelihood $P(\mathbf{d}|r)$ is then given by the value of the conditional pdf, that is
\begin{equation}
    \label{eq:mvn}
    P(\mathbf{d}|r) = \frac{-\frac{1}{2}(\mathbf{d}-\mathbf{\mu}_r)^T\Sigma_r^{-1}(\mathbf{d}-\mathbf{\mu}_r)}{\sqrt{(2\pi)^k |\Sigma_r|}},
\end{equation}
where $|\cdot|$ represents the determinant. The mean $\mathbf{\mu}_r$ and covariance $\Sigma_r$, representing the variability around the cluster centre, are estimated from the conditional distance distributions obtained from the $k$-means clustering results for each regime. These estimates are done separately for ERA-Interim and SEAS5 to avoid biases due to the regimes being slightly different. The estimates of the mean and covariance are surprisingly similar between both datasets, indicating that, apart from the slight difference in regimes, the model does a reasonable job in representing the variability of the regime dynamics. A further discussion on this, including a robustness analysis of the distance distributions is given in the Supplementary Material.

To obtain the prior probability $P(r)$ there is a natural choice from propagating the probabilities of the previous timestep forward. From $k$-means clustering an estimate of the regime dynamics is known, which is characterised by the climatological regime frequencies $P^c$ and transition probabilities $T^c_{ij}$ between the regimes. For SEAS5 these are given by \citep{Falkena2022} (for the regimes ordered as in Figure \ref{fig:regimes_ens})
\begin{equation}
    \label{eq:climatology_probandtrans}
    P^c = 
    \begin{pmatrix}
        0.176 \\
        0.158 \\
        0.160 \\
        0.163 \\ 
        0.175 \\
        0.168
    \end{pmatrix}, \qquad
    T^c = 
    \begin{pmatrix}
        0.728 & 0.000 & 0.039 & 0.062 & 0.060 & 0.112 \\
        0.000 & 0.822 & 0.050 & 0.046 & 0.053 & 0.029 \\
        0.079 & 0.054 & 0.702 & 0.075 & 0.021 & 0.069 \\
        0.069 & 0.058 & 0.065 & 0.739 & 0.037 & 0.031 \\
        0.072 & 0.032 & 0.035 & 0.045 & 0.771 & 0.045 \\
        0.065 & 0.033 & 0.095 & 0.029 & 0.070 & 0.708
    \end{pmatrix}.
\end{equation}
Starting from the regime probabilities at time $t-1$, a best estimate of the prior probabilities for the next time step is 
\begin{equation}
    \label{eq:prior}
    P(t) = T^c P(t-1|\mathbf{d}),
\end{equation}
where $P(t)$ is the vector of prior probabilities $\{P(r)\}_{r=1,...,k}$ at time $t$ and $P(t-1|\mathbf{d})$ the vector of posterior probabilities $\{P(r|\mathbf{d})\}_{r=1,...,k}$ at time $t-1$. Note that in the transition matrix $T^c$ the diagonal elements --- corresponding to persistence of the current regime --- dominate. At the start of each winter, on December 1st, there is no previous probability to use, and thus little prior information on the probability of being in any of the regimes. For that reason the climatological regime frequencies $P^c$ are used as a prior. Note that this is nearly as uninformative as using a uniform distribution.
Here the hard regime assignment is used to obtain both the initial prior for each winter and the transition probabilities to obtain subsequent priors. This is by no means the only option, e.g. one could also use a uniform prior at the start of winter. The choice made here is closest to existing methods and therefore least biased when comparing the results.

Using the prior probabilities $P(r)$ and likelihood of the data $P(\mathbf{d}|r)$ following the conditional distance distributions we can compute the posterior Bayesian probability of a regime given the data $P(r|\mathbf{d})$ using Bayes Theorem \eqref{eq:bayesth} in every timestep. This yields a sequential probabilistic regime assignment, where the regime probabilities of one day are used to obtain a prior for the next day. Applying this method to ERA-Interim data and the ensemble members of the SEAS5 ensemble yields a probability of being in each of the six regimes at every day in winter. From here on we refer to this posterior Bayesian probability simply as the Bayesian probability. 
This Bayesian approach can be related to a HMM approach, where the regime patterns and their transition probabilities are given a priori, leaving only the hidden regime assignment to be discovered. Here the used likelihood differs from that commonly used in the standard Expectation-Maximization (EM) algorithm \citep[e.g.][]{dempster1977,rabiner1989}. In case the transition matrix $T$ cannot be obtained directly, as is done here through observation of the hard regime assignment, one could employ techniques to find $T$ via algorithms designed in the context of HMMs.

The above described sequential Bayesian regime assignment is simple and allows for a straightforward comparison with the commonly used hard regime assignment, as well as with the regularised clustering results (without the need of selecting a constraint parameter). However, there are other options to model the uncertainty and to update the corresponding model parameters sequentially. For instance one can model each regime individually and associate its center estimates with the mean of a Gaussian. The updating procedure for such a model is called the Kalman filter \citep{jdw:Kalman1960} or the corresponding Monte Carlo approximation the Ensemble Kalman Filter \citep{jdw:EvensenLeeuwen2000}, and of course various other methods for more general distributions as well as iterative assimilation of incoming information exist \citep[e.g.][]{Kantas2014,flowfilter,Acevedo2018}. 
The method used here is closer to a particle filter \citep{delMoral1997,Doucet2001} as our ensemble members are weighted with importance weights stemming from the likelihood rather than using an analytic formula such as is used in the Kalman filter.
However, in this paper we specifically aim to stay close to existing methods and model the process of hard regime assignments as random variables in each time step. This allows for a straightforward implementation which can be readily applied in an operational setting. Furthermore, using this method we can investigate whether the results are comparable to those found using regularised clustering methods, which have been used to improve the regime persistence in the identification procedure, without the need to select a constraint parameter.

\subsection{Evaluation}
\label{sec:sequentialbayes_evaluation}

The first question to answer is what the effect is of this Bayesian approach in practice, and whether this matches the intuition behind the method. How does the prior affect the Bayesian probabilities? A next step is to compare the probabilistic approach with results obtained using a hard regime assignment, as given by $k$-means clustering. Is the average regime frequency affected? What is the effect on the regime persistence? In this section we start by discussing the first question by looking at some examples to get a sense for how the method is working in practice, after which we look at the statistics of the results compared to a $k$-means clustering approach to answer the other questions.

\begin{figure}[h]
    \centering
    \includegraphics[width=1.\textwidth]{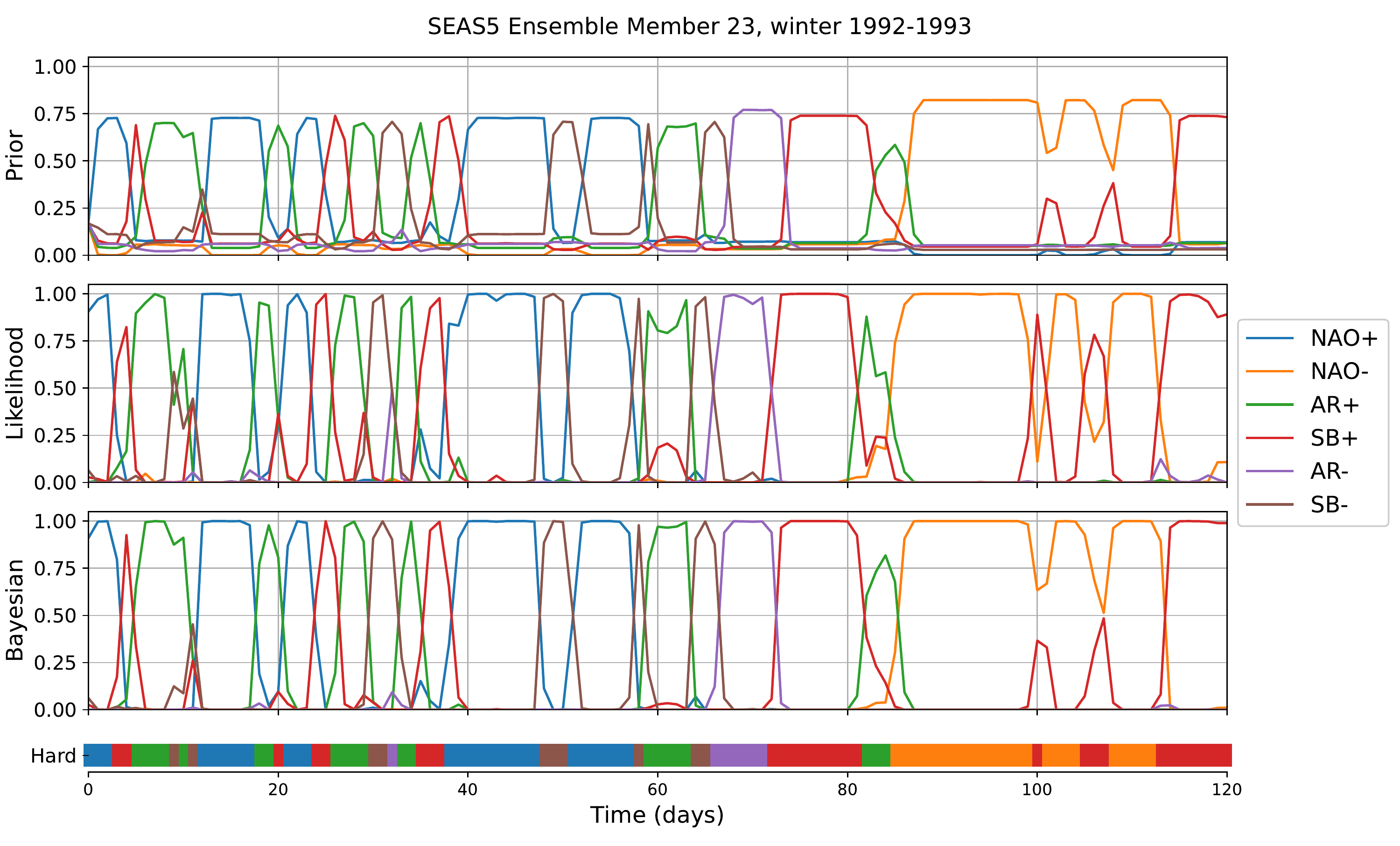}
    \caption{The prior probability, conditional regime likelihood and Bayesian regime probability for the 23rd ensemble member in the sequential Bayesian regime assignment procedure for the winter of 1992-1993. The bar at the bottom indicates the hard regime assignment following $k$-means clustering.}
    \label{fig:regimeocc_compare}
\end{figure}

To start, we consider the Bayesian regime probabilities for a single randomly chosen ensemble member. As the sequential Bayesian regime assignment works on a single-member basis this is the best way to gain insight into the workings of the Bayesian method. In Figure \ref{fig:regimeocc_compare} the prior and Bayesian regime probabilities for the 23rd ensemble member are shown
together with the climatological likelihood corresponding to the observed datapoint. 
A first aspect to note is that most of the time the regime likelihood $P(\mathbf{d}|r)$ gives a clear indication of the regime the data belong to. Secondly, we see that the prior quite closely follows the Bayesian probabilities with a delay of one day, corresponding to the high persistence in the transition matrix (Equation \eqref{eq:climatology_probandtrans}). The initial prior, given by the climatological values, is uninformative and in that case the regime likelihood nearly fully determines the Bayesian probabilities. Subsequently, the prior is much more informative but in most cases the regime likelihood still strongly determines the final probability. However, when the likelihood does not clearly point towards one regime, e.g. around days 8-12, the prior information shifts the probabilities towards stronger persistence, in this case of the AR+ regime. This can also be seen around day 99-101, corresponding to days 7-9 in the example shown in Figure \ref{fig:conceptual} in Section \ref{sec:intro}, where the inclusion of prior information favors persistence over a short excursion away from the most likely regime. In this way the Bayesian regime assignment allows for identifying stronger persistence, i.e. high probability of the dominant regime, without losing the signal of other regimes entering the dynamics as they still have some non-zero probability. The effect of this approach for ERA-Interim data is similar.

The Bayesian probabilistic regime assignment allows to understand some of the subtleties of the regime dynamics, e.g. regime transitions occur in the form of a decrease/increase of the regime probabilities. How does such an approach compare to the commonly used hard regime assignment obtained using $k$-means clustering? The bar at the bottom of Figure \ref{fig:regimeocc_compare} shows the hard regime assignment corresponding to this time series. The Bayesian regime probabilities vary more smoothly, and show less short back-and-forth transitions between regimes which occur several times for the hard regime assignment, e.g. around day 9 and 20. 
In \citet{Falkena2020} a constraint on the number of transitions between regimes was introduced to reduce the number of short back-and-forth transitions between regimes, based on the regularised clustering method introduced by \citet{Horenko2010a}. This was shown to increase the regime persistence without affecting the regime occurrence rates, provided the constraint parameter was chosen appropriately.
The optimal constraint parameter corresponded to an average regime duration of 6.3 days. It was selected by considering the Bayesian Information Criterion and falls within the region where the regime occurrence rates are not affected by the regularisation.

\begin{figure}[h]
    \centering
    \includegraphics[width=1.\textwidth]{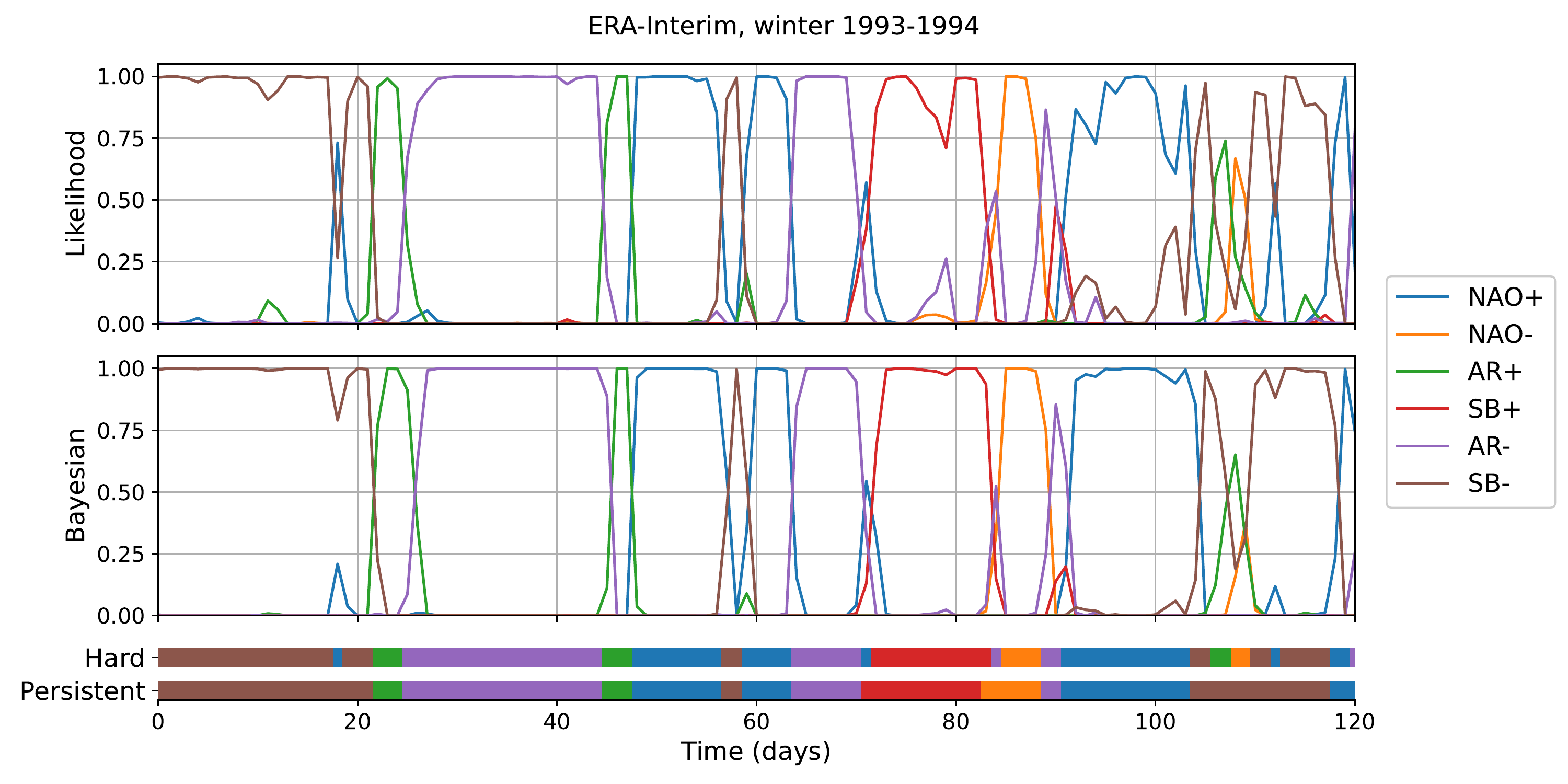}
    \caption{The observed regime likelihood and Bayesian regime probability for ERA-Interim, with the hard assignment using a standard or time-regularised (persistent) $k$-means algorithm shown by the bars for the winter of 1993-1994.}
    \label{fig:regimeocc_eraicompare}
\end{figure}

In Figure \ref{fig:regimeocc_eraicompare} a comparison between the regime likelihood, Bayesian regime probabilities and a hard regime assignment obtained using either a standard or regularised $k$-means approach is shown for ERA-Interim for the winter of 1993-1994. The regularisation does reduce the number of regime transitions, by e.g. removing the NAO+ regime between two occurrence of SB$-$ around day 18. At the same time the Bayesian probabilities show a small increase in the NAO+ likelihood, with SB$-$ still having the highest probability. Here the regularisation and Bayesian approach thus yield similar results. On the other hand, around e.g. day 84 and 107 the regularisation eliminates some regime transitions where the Bayesian probabilities still show some signal of the corresponding regimes. The probabilistic approach thus allows to identify the data where the regime assignment is less clear, showing an increase in probability instead of a hard regime change. It also retains some regime transitions that the regularised clustering eliminates due to it being difficult to select the ``correct'' constraint value. In the probabilistic approach these show as increases in the corresponding regime probability. This analysis confirms that the Bayesian approach seems to be doing something sensible, without having to tune any parameters.
When the data clearly belongs to one of the six regimes, there is little benefit to the Bayesian approach. The main times where it makes a difference are the periods when one regime transitions into another, or when a regime loses some of its strength in favor of another regime but then gains in strength again. Such a reduction in the regime probabilities could be an indication of increased flow instability, being close to transitioning into another of the six canonical states.

\begin{figure}[h!]
    \centering
    \includegraphics[width=1.\textwidth]{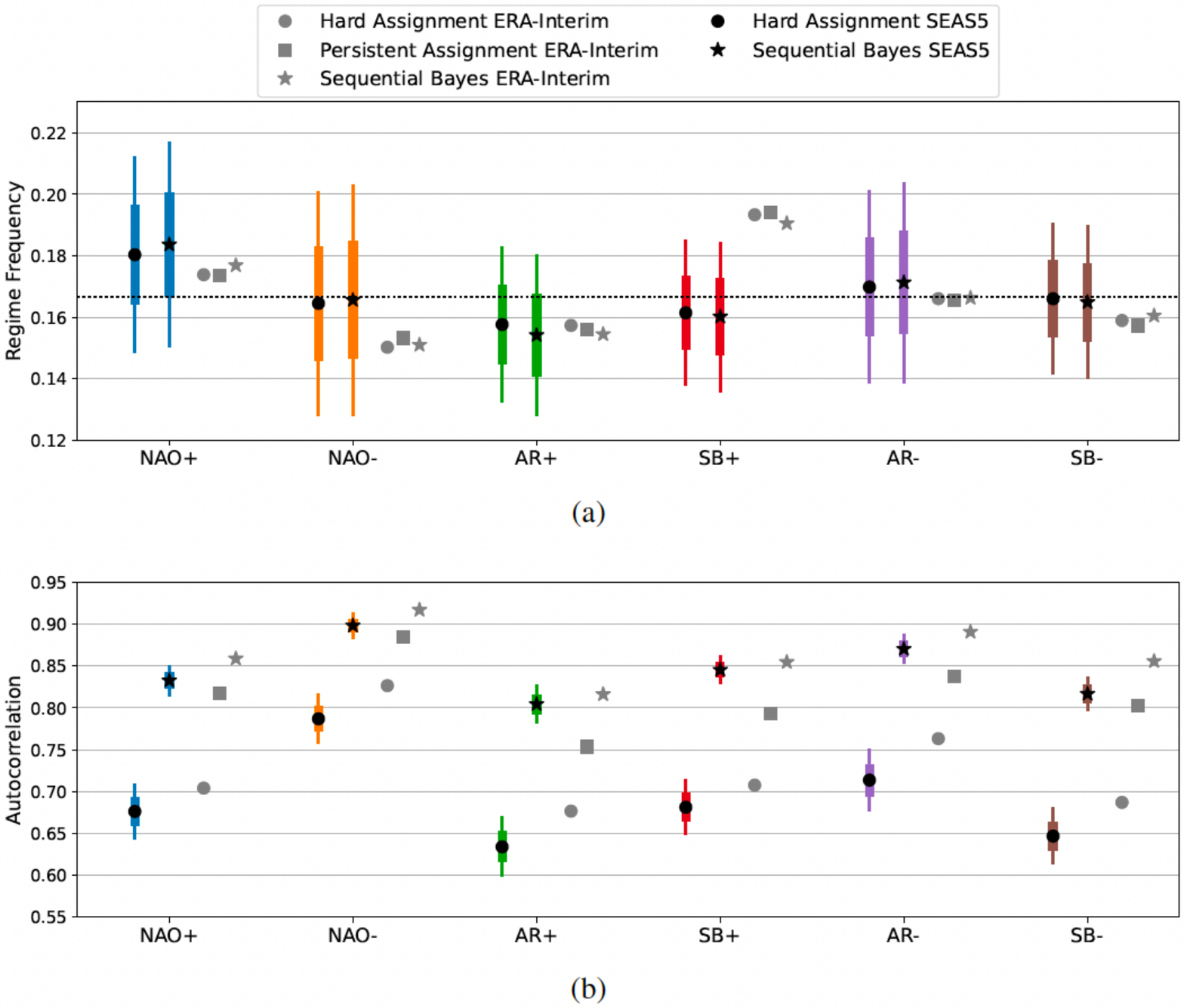}
    \caption{The regime frequencies and 1-day autocorrelation as obtained using either standard $k$-means clustering (circles) or a sequential Bayesian regime assignment (stars) for the SEAS5 hindcast ensemble (symbols with error bars) and ERA-Interim (symbols only), for which also the values obtained with the time-regularised $k$-means clustering method are shown (squares). Error bounds are determined using bootstrapping with one member per year (with replacement, 500 times), where the thick bars indicate the plus-minus one standard error range with thin bars extending showing the 95\% confidence interval.}
    \label{fig:compare_sequential}
\end{figure}

The impact of the sequential Bayesian approach on the regime frequencies, computed as the average Bayesian regime probability for this method, and (1-day) autocorrelation is shown in Figure \ref{fig:compare_sequential}. 
Here the autocorrelation for the hard regime assignment is computed using a time series which is one when data is assigned to the corresponding regime and zero otherwise.
The average frequencies of the regimes do not change when using the Bayesian regime assignment, as can be seen in Figure \ref{fig:compare_sequential}(a). This holds both for the SEAS5 hindcast ensemble data and for ERA-Interim, where also the results of the regularised $k$-means clustering algorithm are shown for comparison. On the other hand the autocorrelation, being an indication of the persistence of the regimes, is strongly affected (Figure \ref{fig:compare_sequential}(b)). For ERA-Interim we see that the sequential Bayesian approach increases the autocorrelation even beyond that obtained using a regularised clustering algorithm that contains a persistence constraint. Also for SEAS5 a strong increase in autocorrelation is found using the sequential Bayesian regime assignment compared to a standard hard assignment. For most regimes the ERA-Interim values lie at the top of the SEAS5 autocorrelation range, both for the standard and Bayesian approach. Thus we find that the Bayesian approach does not alter the regime frequencies, but does lead to more persistent regime dynamics, as we might hope. 
This suggests that the transition probabilities in Equation \eqref{eq:climatology_probandtrans}, which are used to obtain the prior regime probabilities, likely are an underestimation of the true persistence, which is improved by the use of Bayes Theorem.

% \begin{figure}[h!]
%     \centering
%     \begin{subfigure}{1.\textwidth}
%         \centering
%         \includegraphics[width=1.\textwidth]{Occ_boxplot_k6.eps}
%         \caption{}
%         \label{fig:occ_compare}
%     \end{subfigure}
%     \begin{subfigure}{1.\textwidth}
%         \centering
%         \includegraphics[width=1.\textwidth]{Autocor_boxplot_k6.eps}
%         \caption{}
%         \label{fig:autocor_compare}
%     \end{subfigure}
%     \caption{The regime frequencies and 1-day autocorrelation as obtained using either standard $k$-means clustering or a sequential Bayesian regime assignment for the SEAS5 hindcast ensemble (boxes) and ERA-Interim (circles and stars), for which also the values obtained with the time-regularised $k$-means clustering method are shown (squares). Error bounds are determined using bootstrapping with one member a year (with replacement, 500 times), where the boxes indicate the interquartile range with the whiskers extending 1.5 times beyond that and the circles being outlier points.}
%     \label{fig:compare_sequential}
% \end{figure}

% \begin{figure}[h!]
%     \centering
%     \begin{subfigure}{1.\textwidth}
%         \centering
%         \includegraphics[width=1.\textwidth]{Occ_confintplot_k6.pdf}
%         \caption{\Huge{}}
%         \label{fig:occ_compare}
%     \end{subfigure}
%     \begin{subfigure}{1.\textwidth}
%         \centering
%         \includegraphics[width=1.\textwidth]{Autocor_confintplot_k6.pdf}
%         \caption{\Huge{}}
%         \label{fig:autocor_compare}
%     \end{subfigure}
%     \caption{The regime frequencies and 1-day autocorrelation as obtained using either standard $k$-means clustering (circles) or a sequential Bayesian regime assignment (stars) for the SEAS5 hindcast ensemble (symbols with error bars) and ERA-Interim (symbols only), for which also the values obtained with the time-regularised $k$-means clustering method are shown (squares). Error bounds are determined using bootstrapping with one member per year (with replacement, 500 times), where the thick bars indicate the one-standard error range with thin bars extending showing the 95\% confidence interval.}
%     \label{fig:compare_sequential}
% \end{figure}

\section{Ensemble Bayesian Regime Assignment}
\label{sec:ensemblebayes}

The implicit assumption made in the sequential Bayesian approach as discussed in the previous section is that the regime dynamics is statistically stationary in time. That is, the climatological likelihood functions and transition probabilities do not change in time. This is a reasonable and minimal first assumption yielding good results, but it is likely that external factors such as ENSO affect some aspects of the regime dynamics as discussed in Section \ref{sec:intro}. 
There are two obvious ways in which to include the effect of external forcing in the Bayesian approach. The first is to update the regime likelihood functions in time. The second is to update the prior probabilities. These two aspects are by no means the only aspects of the regime dynamics that can be affected by external forcing. For example, one can imagine that the regimes themselves change as a consequence of external factors causing changes in the climate system. However, this is nearly impossible to quantify with the limited available data and no robust evidence for this has been found so far \citep[e.g.][]{Corti1999,Dorrington2022}. Therefore, we only discuss the above-mentioned two approaches.

In the following analysis we focus on the latter of these two approaches. The main reason for this is the lack of data availability. Even though the SEAS5 hindcast ensemble has 51 members for each year, this still is insufficient to allow for e.g. weekly updating of the likelihood functions. An option for which sufficient data are available would be to compute the likelihood function during e.g. strong El Ni\~no years, and use those to change the likelihood functions each year. However, this relies on the hypothesis that the regions in phase space belonging to each of the regimes shift as a consequence of ENSO forcing, while it may simply be the case that some regions are visited more often than others. As there are only 36 years of data available it is impossible to test this hypothesis and thus we refrain from pursuing this approach further. On the other hand, there is sufficient data to update the prior probabilities in time. There are several ways in which this can be done. For example, one can use information on ENSO to shift the prior probabilities, or one can make use of the ensemble information by allowing the transition probabilities to change in time. We pursue the latter approach, as it makes use of the information within the SEAS5 ensemble and does not require any external information. It is explained and evaluated in the next two sections followed by an analysis of the resulting interannual variability in Section \ref{sec:interannual}. 

\subsection{Updating the Transition Probabilities}
\label{sec:ensemblebayes_transition}

To obtain more informative prior regime probabilities, the transition probabilities $T_{ij}$ from regime $i$ to $j$ are updated following the ensemble behavior. This allows not only for (fixed) persistence to inform the prior, but also non-stationary external factors, such as ENSO, through the ensemble statistics. Although there is not sufficient data to robustly estimate the transition probabilities directly, they can be inferred from the occurrence rates. The main assumption we make when updating the transition matrix $T$ in time is that the regime probabilities are approximately stationary with respect to the current best estimate of the transition matrix. That is, we look for a transition matrix $T(t)$ for which the regime probabilities averaged over the ensemble at time $t$, $\bar{P}(t)$, are approximately stationary:
\begin{equation}
    \label{eq:trans_stat}
    T(t) \bar{P}(t) = \bar{P}(t) + \epsilon^t.
\end{equation}
Here $\epsilon^t$ is a noise term. Note that the climatological transition probabilities $P^c$ are (nearly) stationary with respect to the transition matrix $T^c$. The aim thus is to find a transition matrix $T(t)$ for which Equation \eqref{eq:trans_stat} holds. In addition we have that a transition matrix is normalised, meaning its columns each sum to unity:
\begin{equation}
    \label{eq:trans_norm}
    \sum_{i=1}^k T_{ij} = 1, \qquad \forall j \in {1,...,k}.
\end{equation}
This gives two equations which are used to update $T(t)$ at each timestep $t$.
The problem of finding the values of the transition matrix $T(t)$ is ill-posed as there are not sufficient constraints, which means some choices need to be made in determining its values. The approach we propose in the following paragraph is one that follows the regime dynamics closely and is least biased in the sense that the deviations from $T^c$ are equally distributed over all six regimes.

The regime dynamics is dominated by persistence, i.e. the probability of a regime to transition to itself corresponding to the diagonal elements of the transition matrix, as can be seen in Equation \eqref{eq:climatology_probandtrans}. Therefore we focus on these diagonal elements $T_{ii}(t)$ for updating the matrix $T(t)$ in time. Writing out Equation \eqref{eq:trans_stat} elementwise while separating the diagonal and off-diagonal elements yields
\begin{equation}
    T_{ii}(t) \bar{P}_i(t) + \sum_{j \neq i}^k T_{ij}(t) \bar{P}_j(t) = \bar{P}_i(t) + \epsilon^t_i, \qquad \forall i \in {1,...,k}.
\end{equation}
As the diagonal terms dominate, we assume the off-diagonal elements do not differ much from the climatological values, that is $T_{ij}(t) \approx T_{ij}^c$ for all $i\neq j$. This yields an approximate equation for the diagonal elements of $T(t)$:
\begin{equation}
    \label{eq:trans_update}
    T_{ii}(t) \bar{P}_i(t) \approx \bar{P}_i(t) - \sum_{j \neq i}^k T_{ij}^c(t) \bar{P}_j(t).
\end{equation}
When a particular regime is less populated than it is in climatology, the other regimes will conversely be more populated, implying a larger negative term on the right-hand side of \eqref{eq:trans_update} and thus a smaller value of the self-transition probability, which makes physical sense. Note that this approximation breaks down when $\bar{P}_i(t)$ is very small compared to the other $\bar{P}_j(t)$, in which case we set $T_{ii}(t) = 0$ to prevent negative values. Starting from the updated diagonal elements, the off-diagonal elements are computed using Equation \eqref{eq:trans_norm} with an equal distribution of the perturbation from the climatological value over the off-diagonal terms. 

The estimation of the transition matrix $T$ in essence is the same as trying to fit a HMM to the data. The difficulty here is the limited availability of data, where we only consider data at one point in time to retain the sequential nature of the method. This makes the use of less heuristic, more sophisticated methods unreliable due to the large impact of noise on the data. If many more ensemble members would be available, something like the Baum-Welch algorithm might be a worthwhile approach for estimating $T$ \citep{baum1970}.
Starting the updating of $T(t)$ from the diagonal elements and adjusting the off-diagonal elements equally is not the only option. It might even be better to not adjust the off-diagonal elements equally. However, since $\bar{P}(t)$ is an average over only 51 ensemble members, robustness would be an issue when making any further assumptions in updating $T(t)$ and hence we stick to the simplest approach.

The above method is equivalent to considering $T(t)$ as the climatological transition matrix plus a perturbation, i.e. $T(t) = T^c + T'(t)$, and subsequently assuming that the perturbations to the off-diagonal terms are small. An alternative way of looking at this is by considering it as a Markov regression model \citep{Hamilton1989,Krolzig1997}. That is, we write the transition matrix $T$ as
\begin{equation}
    \label{eq:trans_markov}
    T(t) = T^c + \sum_m \alpha_m(t) T_m.
\end{equation}
Here $T_m$ are matrices that set the shape of the perturbations to the climatological transition matrix, where the sum over each of the columns is zero for every $m$, and $\alpha_m(t)$ gives the strength of that term at time $t$. For a choice of
\begin{equation}
    T_m = 
    \begin{pmatrix}
        0 & \dots & -\frac{1}{k-1} & \dots & 0 \\
         & & \vdots & &  \\
        \vdots & & 1 & & \vdots \\
         & & \vdots & &  \\
        0 & & -\frac{1}{k-1} & & 0 \\
    \end{pmatrix},
\end{equation}
where the $m$-th column is non-zero this is exactly equivalent to the approach mentioned before. Here the $\alpha_m$ can be computed using the same assumptions as discussed before. This shows that there are several ways of looking at the problem that yield the same outcome, increasing the confidence in this approach.

\subsection{Evaluation}
\label{sec:ensemblebayes_evaluation}

\begin{figure}[h]
    \centering
    \includegraphics[width=1.\textwidth]{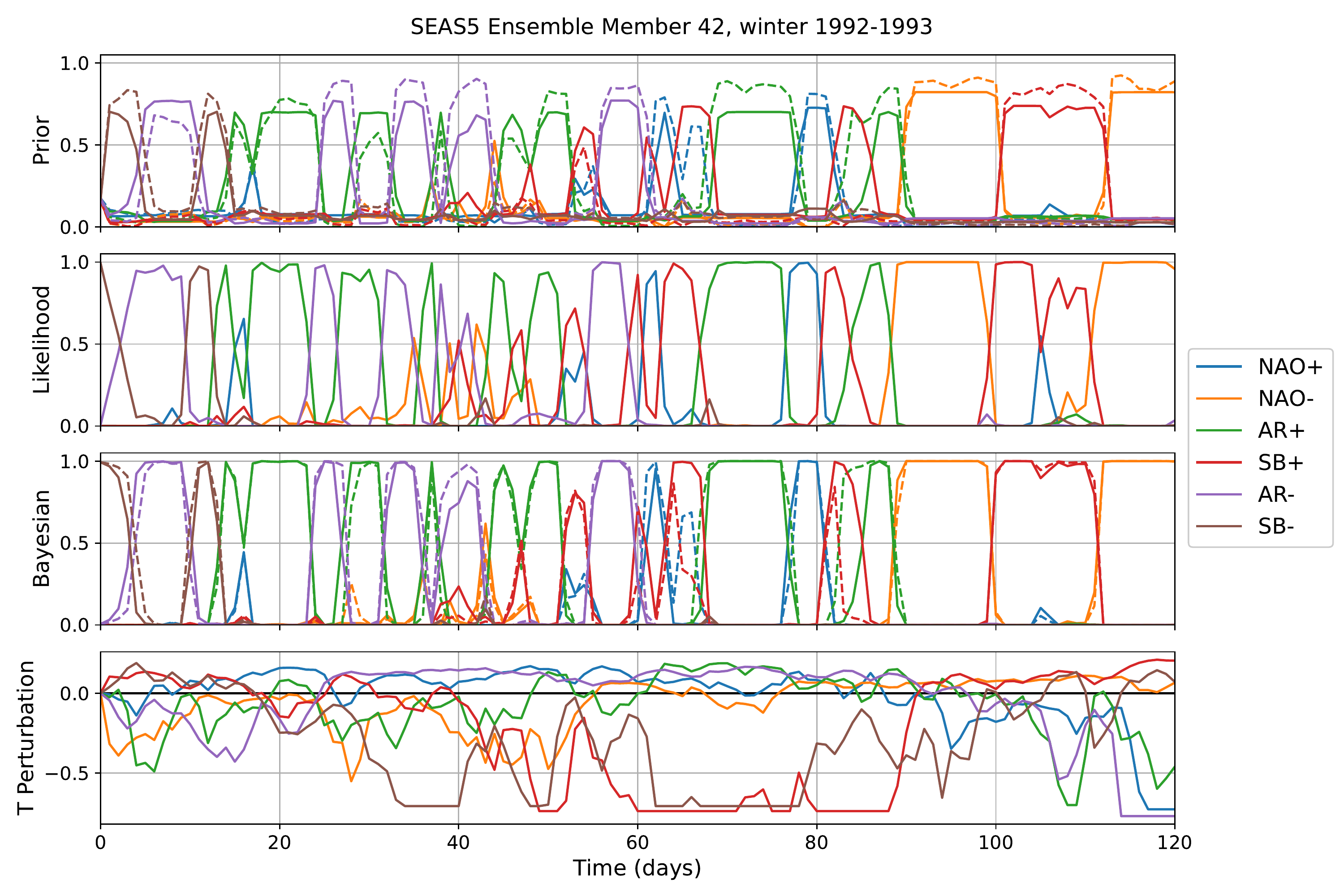}
    \caption{In the top three panels the prior probability, conditional regime likelihood and Bayesian regime probability for the 42nd ensemble member in the Bayesian regime assignment procedure for the winter of 1992-1993 are shown. The solid line shows the sequential Bayesian approach and the dashed line the ensemble approach discussed in this section. The bottom panel shows the difference between the updated self-transition probabilities in the ensemble approach and the climatological values.}
    \label{fig:regimeocc_compare_inclens}
\end{figure}

To get an idea of how this approach can inform the prior probabilities consider Figure \ref{fig:regimeocc_compare_inclens}, which shows both the sequential and ensemble Bayesian regime assignments for the (randomly chosen) 42nd ensemble member during the winter of 1992-93. This is the same winter for which the 23rd ensemble member is shown in Figure \ref{fig:regimeocc_compare}. As an example, consider the probability of AR$-$. Around days 5-10 the ensemble indicates this regime is less likely, as shown by a lower self-transition probability, lowering the prior probability of the regime. On the other hand, from day 25 onward AR$-$ is more likely according to the ensemble, increasing its prior probability compared to the sequential approach. In most cases changes to the final probabilities are small. The only exceptions occur when a regime is deemed very unlikely, i.e. does not occur in any of the other ensemble members, as happens twice for the SB+ regime between day 60 and 90. In these two cases a high observed likelihood for SB+ is reduced substantially in the Bayesian probabilities in favor of the second most-likely regime according to the likelihood, e.g. a 90\% likelihood is reduced to a 35\% Bayesian probability. Yet importantly, the Bayesian probability of this regime is still non-zero, so it can quickly respond to new information. The overall regime frequencies and autocorrelation are not affected and remain as shown in Figure \ref{fig:compare_sequential} for the sequential approach.

\section{Interannual Variability}
\label{sec:interannual}

\begin{figure}
    \centering
    \includegraphics[width=1.\textwidth]{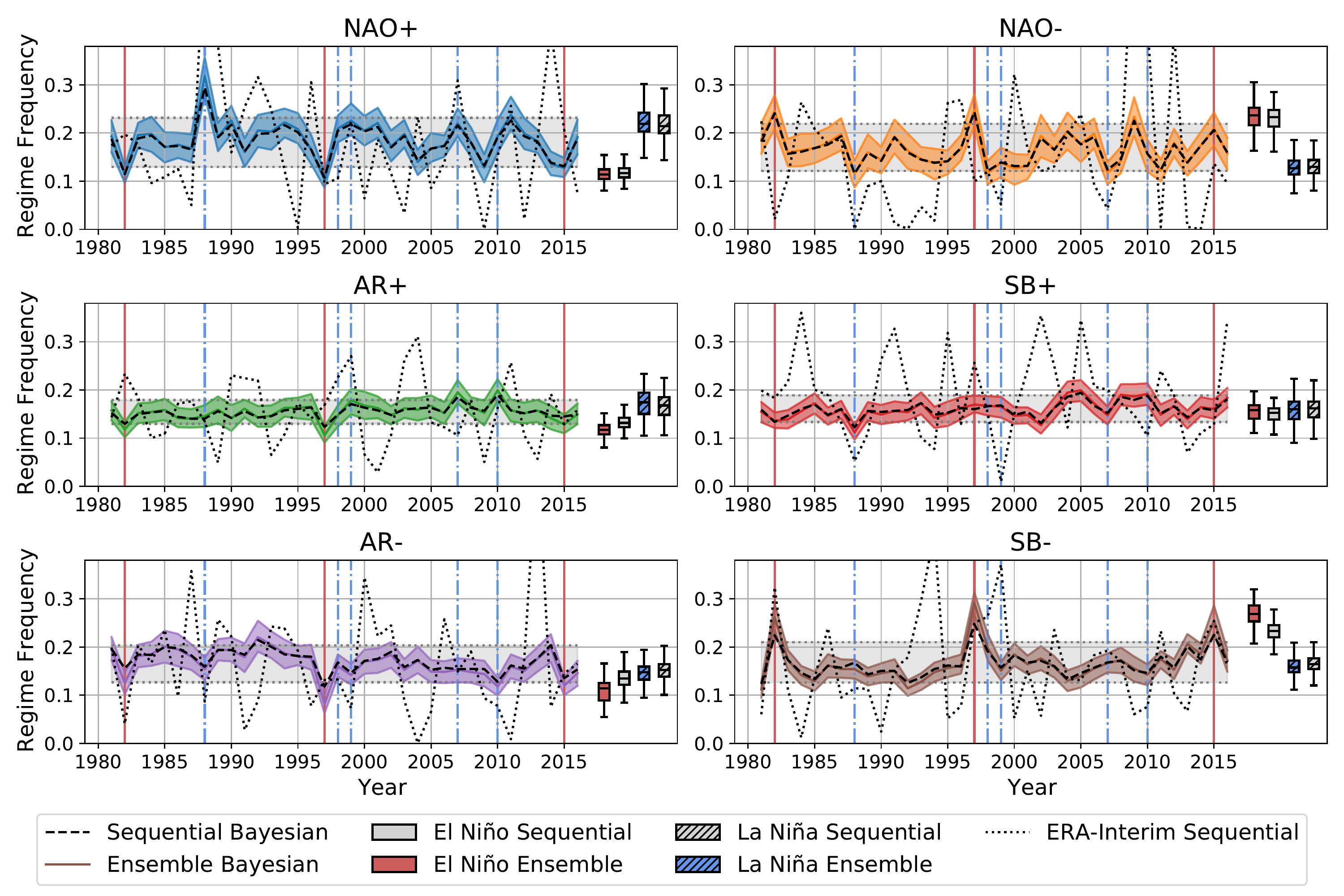}
    \caption{The interannual variability of the occurrence rates for the ensemble Bayesian regime assignment for SEAS5 (color, with 95\% confidence interval shaded), with the sequential Bayesian approach indicated by the black dashed lines. The grey shaded areas bounded by the grey dotted lines indicate the 10th and 90th percentile of the ensemble Bayesian assignment for each regime. The black dotted curve shows the ERA-Interim variability and the box-and-whisker plots on the right show the average occurrence rate during very strong El Ni\~no (indicated by the vertical red solid lines) and strong La Ni\~na years (indicated by the vertical blue dash-dotted lines).}
    \label{fig:occrate_yearcomp}
\end{figure}

The interannual variability as obtained using the ensemble Bayesian regime assignment is shown in Figure \ref{fig:occrate_yearcomp}, with the result of the sequential Bayesian approach shown for reference (the interannual variability of the sequential Bayesian approach is nearly identical to that obtained for the $k$-means clustering assignment). The primary signal in the variability is found during very strong El Ni\~no years (vertical red solid lines) with SB$-$ and NAO$-$ showing an increase in frequency, while AR+, AR$-$ and NAO+ show a decrease in frequency. 
The signal during strong La Ni\~na years (vertical blue dash-dotted lines) is less clear, 
with on average an increase in NAO+ and decrease of NAO$-$ frequency. However, not every individual event matches this behavior.
To define El Ni\~no and La Ni\~na years the Ni\~no 3.4 index is used \citep{Trenberth1997}. Strong years correspond to a threshold of $\pm1.5$, and very strong years to a threshold of $\pm2$. The asymmetry in the thresholds used for El Ni\~no and La Ni\~na years is due to there being no very strong La Ni\~na events in the considered time period.
These results,
with a less pronounced regime response to La Ni\~na compared to El Ni\~no,
reflect the well-known nonlinearity of the response to ENSO \citep{Straus2004, Toniazzo2006} and are in line with those obtained in \citet{Falkena2022} using a regularisation on the ensemble members.
The boxes on the right of each panel show the average regime frequencies during the identified El Ni\~no and La Ni\~na years for both the sequential and ensemble Bayesian approach, where there is an asymmetric response to ENSO for both methods. Some enhancement of the signal is found using the ensemble Bayesian regime assignment, which is most clear for the AR$-$ and SB$-$ regimes. The ERA-Interim variability from the sequential Bayesian approach is shown as well to give a perspective on the magnitude of the interannual variability. 

\begin{figure}
    \centering
    \includegraphics[width=1.\textwidth]{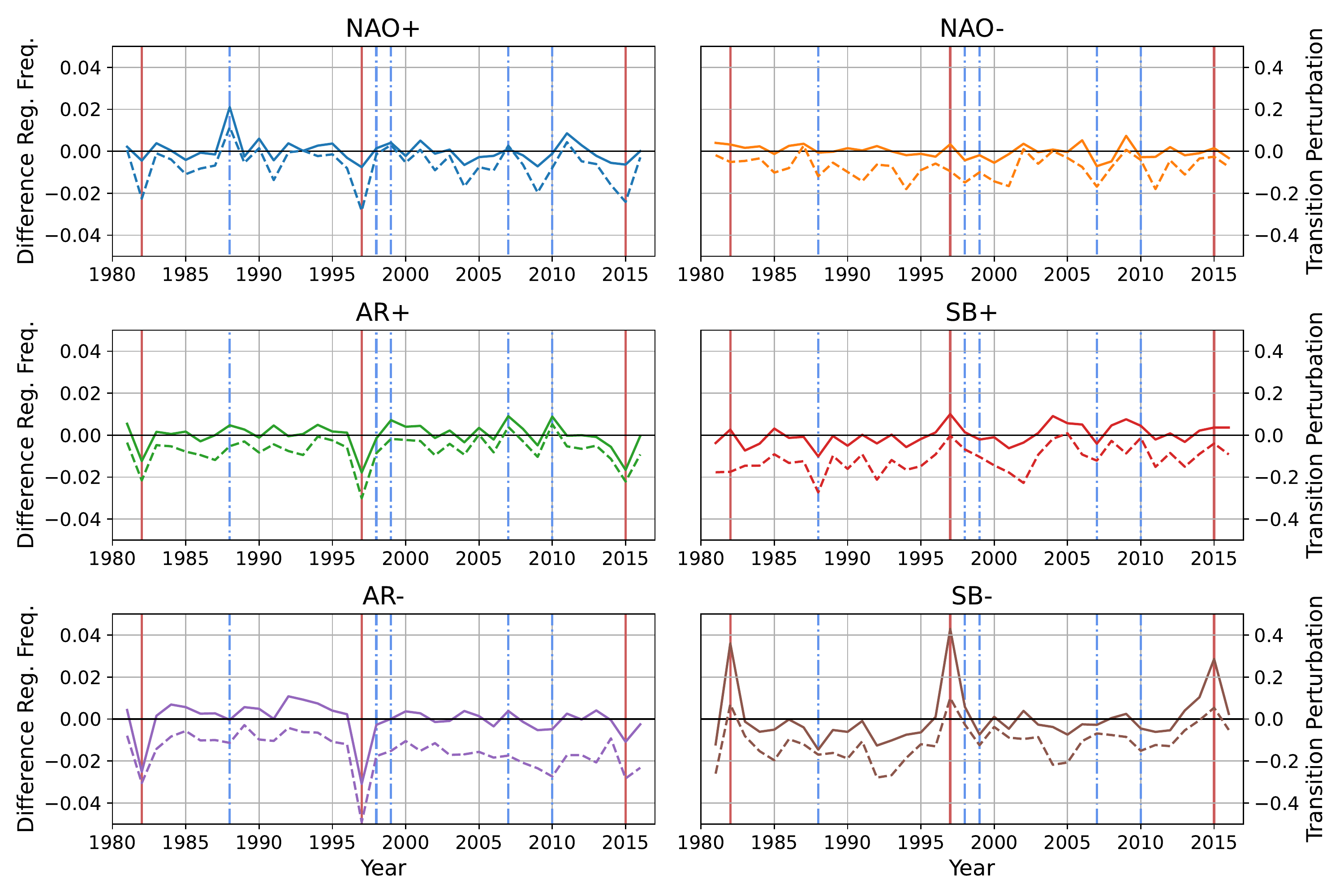}
    \caption{The difference in interannual variability of the occurrence rates between the standard and ensemble Bayesian regime probabilities (solid), as well as the change in the self-transition probability for the regimes following the ensemble (dashed).}
    \label{fig:occrate_yeardiff}
\end{figure}

To further consider the effect the updating of the transition matrix in the ensemble approach has on the interannual variability, consider Figure \ref{fig:occrate_yeardiff} which shows the difference between the sequential and ensemble Bayesian regime assignment as well as the yearly average change to the self-transition probabilities, or persistence, of the regimes following the ensemble approach. Note that on average the perturbation to the self-transition probabilities is negative. The effect of the ensemble Bayesian approach on the regime frequencies is clearly visible for AR+, AR$-$ and SB$-$, where the signal in response to El Ni\~no is enhanced. For NAO$+$ a strong increase in regime frequency is found for the 1988-1989 La Ni\~na, together with a weak change during El Ni\~no years. NAO$-$ and SB+ do not show much difference in interannual variability between the two methods, although in the latter case there is little signal to enhance. The changes in the self-transition probabilities in general match those found in the regime frequencies, as expected. One aspect to note here is that for NAO+ the changes in the self-transition probability are relatively larger than those in the regime frequencies, especially when comparing to SB$-$.

The response of the changes in regime frequency to El Ni\~no events found using the ensemble Bayesian approach appears to show a true signal and is very unlikely to have arisen by chance. To understand this, consider the change in regime frequency for SB$-$. The marginal probability of a very strong El Ni\~no event is 3/36 (3 events in 36 years), so the chance of the first increase in SB$-$ frequency aligning with El Ni\~no is 3/36. Then, given the first El Ni\~no event has already happened, the probability of the second spike aligning is 2/35 and for the third 1/34. This gives a probability of $3/36 \cdot 2/35 \cdot 1/34 \approx 10^{-4}$ for the alignment occurring by chance. The alignment of the increase/decrease in frequency for the other regimes only further decreases the probability of this being by chance. Also note that the response of both AR+ and AR$-$ is a decrease in regime frequency during El Ni\~no years, indicating another aspect of nonlinearity in the circulation response to ENSO.

\begin{figure}
    \centering
    \includegraphics[width=1.\textwidth]{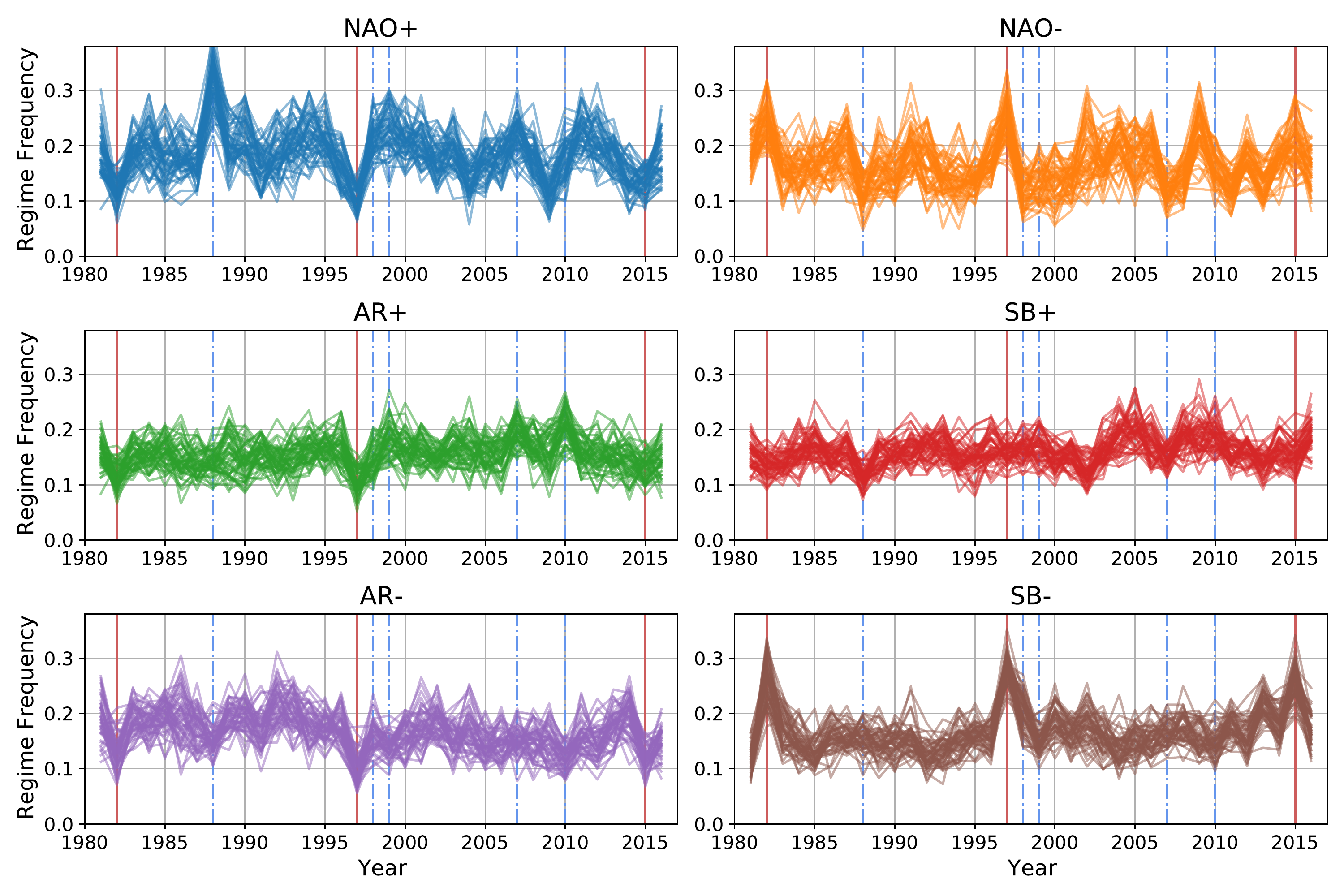}
    \caption{The interannual variability of the regime frequency for the ensemble Bayesian approach when applied to (random) ensembles of 10 members. In total 50 random ensembles are shown. The solid red and dash-dotted blue lines indicate very strong El Ni\~no and strong La Ni\~na years respectively.}
    \label{fig:occrate_10members}
\end{figure}

\begin{figure}
    \centering
    \includegraphics[width=.8\textwidth]{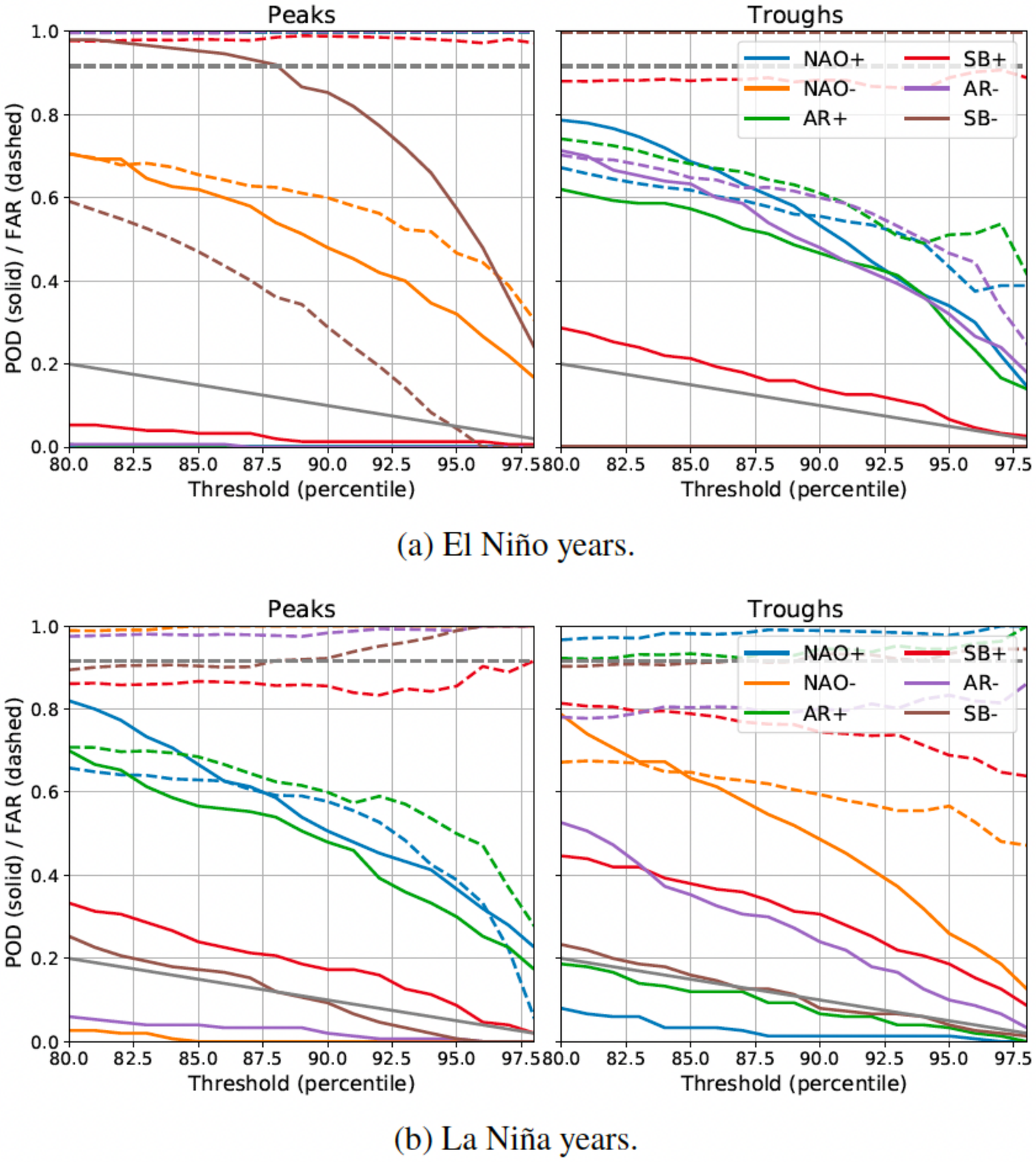}
    \caption{The probability of detection (solid) and false alarm ratio (dashed) for a peak or trough in regime frequency in 10-member subsamples of the SEAS5 ensembles occurring in the same year as a very strong El Ni\~no or strong La Ni\~na, as a function of the percentile used for the definition of the peaks and troughs. The colored lines indicate the regime values, and the grey lines the values for peaks and troughs occurring in random years, i.e. no signal.}
    \label{fig:podfar_enso}
\end{figure}

Some of these signals in response to ENSO can already be picked up using 10-member ensembles. In Figure \ref{fig:occrate_10members} the interannual variability of the regime frequency is shown for 50 random 10-member ensembles obtained from the full SEAS5 ensemble. For the full ensemble the strongest signal was found for SB$-$ during very strong El Ni\~no years, and this is the signal that jumps out most strongly again. To quantify this the Probability of Detection (POD) and False Alarm Ratio (FAR) for the 10-member ensembles are considered for peaks or troughs in regime frequency aligning with El Ni\~no and La Ni\~na (Figure \ref{fig:podfar_enso}). Here, peaks and troughs are considered as exceedances with respect to the $n$th percentile. The POD is computed as the number of peaks/troughs aligning with El Ni\~no/La Ni\~na years over the total number of El Ni\~no/La Ni\~na years, and the FAR is computed as the number of peaks/troughs outside those El Ni\~no/La Ni\~na years divided by the total number of peaks/troughs. As expected, for El Ni\~no there is a high POD for peaks in the SB$-$ regime frequency with a relatively low FAR (Figure \ref{fig:podfar_enso}(a)). Also for NAO$-$ (peaks), NAO+, AR+ and AR$-$ (troughs) there is some signal, with the FAR being comparable to the POD. For La Ni\~na years there is some signal for NAO+, AR+ (peaks) and NAO$-$ (troughs), but it is not as strong as for SB$-$ in El Ni\~no years (Figure \ref{fig:podfar_enso}(b)). This is to be expected as we cannot expect to identify strong signals using a smaller ensemble if they are not clear in the full ensemble. Nevertheless, the relatively high PODs for these three regimes are encouraging.

% \begin{figure}
%     \centering
%     \begin{subfigure}{1.\textwidth}
%         \centering
%         \includegraphics[width=.8\textwidth]{Pod_far_elnino_10test.pdf}
%         \caption{El Ni\~no years.}
%         \label{fig:podfar_elnino}
%     \end{subfigure}
%     \begin{subfigure}{1.\textwidth}
%         \centering
%         \includegraphics[width=.8\textwidth]{Pod_far_lanina_10test.pdf}
%         \caption{La Ni\~na years.}
%         \label{fig:podfar_lanina}
%     \end{subfigure}
%     \caption{The probability of detection (solid) and false alarm ratio (dashed) for a peak or trough in regime frequency in 10-member subsamples of the SEAS5 ensembles occurring in the same year as a very strong El Ni\~no or La Ni\~na, as a function of the percentile used for the definition of the peaks and troughs. The colored lines indicate the regime values, and the grey lines the values for peaks and troughs occurring in random years, i.e. no signal.}
%     \label{fig:podfar_enso}
% \end{figure}

To see whether the found response to ENSO for some regimes also reflects a predictable signal in the observations we regress the ERA-Interim interannual variability onto the SEAS5 one, as in \citet{Falkena2022}. The results for this, looking at the sequential and ensemble Bayesian approach, are shown in Table \ref{tab:regression}. In addition to the $p$-value, we also compute the Bayes factor which is the ratio of the probabilities of the data given two different hypotheses $H_1$ and $H_2$, i.e. $P(D|H_1)/P(D|H_2)$ \citep{Kass1995}. Here the first hypothesis $H_1$ is that of a linear regression model, whereas the second hypothesis $H_2$ assumes a constant, climatological, regime frequency. For its computation we follow the Bayesian Information Criterion approximation from \citet{Wagenmakers2007}. Values of the Bayes factor above one indicate $H_1$ is more likely, with values between 3 and 20 constituting positive evidence and values over 20 yielding strong evidence towards it \citep{Kass1995}.

Using the sequential Bayesian approach we already find some predictable signal for the NAO+ and SB$-$ regimes, with Bayes factors of 7.6 and 5.1 respectively (Table \ref{tab:regression}). The Bayes factor for NAO$-$ is also above 3, but here the $p$-value is larger reducing the confidence in this being a true signal. These results are comparable with those found in \citet{Falkena2022}, with the regression coefficients being close to one for NAO+, NAO$-$ and SB$-$. These regression coefficients around one indicate the signal in SEAS5 is of similar magnitude to that in ERA-Interim, showing no evidence of a signal-to-noise paradox for the regime frequencies, in contrast to the NAO-index \citep{Falkena2022}. Using the ensemble information to update the transition probabilities increases the predictable signal for NAO+ and SB$-$, with smaller $p$-values and higher Bayes factors. Also the AR$-$ signal is enhanced with a Bayes factor over 3 although the $p$-value is still relatively large. The enhancement of the NAO+ signal is comparable to that found using a regularised clustering approach, whereas the change for SB$-$ is weaker (a Bayes factor of 13.2 compared to 5.5, \citet{Falkena2022}). On the other hand, the decrease in Bayes factors for NAO$-$ and AR$-$ using a regularised approach is not found using the ensemble Bayesian method, which shows small increases of the Bayes factors. In \citet{Falkena2022} a significant signal was found using multiple linear regression of ERA-Interim NAO$-$ onto the SEAS5 NAO+ and SB$-$, which we find here as well with Bayes factors of 21.1 for the sequential method increasing to 26.6 using the ensemble approach. Comparing the two methods, we find that the ensemble Bayesian regime assignment allows to identify more pronounced interannual variability signals for some regimes while still accounting for the signal of the other regimes.

\begin{table}[]
    \centering
    \begin{tabular}{cc|cccccc|cc}
         & Regime & NAO+ & NAO$-$ & AR+ & SB+ & AR$-$ & SB$-$ & MLR & NAO$-$ \\
         \hline
        Sequential Bayes &  Reg. Coeff. & 1.170 & 1.094 & -0.504 & 0.258 & 1.207 & 1.083 & NAO+ & -1.369 \\
         & & & & & & & & SB$-$ & -1.838 \\
         & p-value & 0.052 & 0.139 & 0.592 & 0.795 & 0.174 & 0.082 & & 0.047 \\
         & Bayes Fac. & 7.579 & 3.251 & 1.167 & 1.037 & 2.696 & 5.054 & & 21.108 \\
        \hline
        Ensemble Bayes &  Reg. Coeff. & 1.066 & 1.035 & -0.435 & 0.225 & 1.037 & 0.785 & NAO+ & -1.429 \\
         & & & & & & & & SB$-$ & -1.412 \\
         & p-value & 0.044 & 0.133 & 0.527 & 0.782 & 0.136 & 0.075 & & 0.041\\
         & Bayes Fac. & 8.910 & 3.365 & 1.240 & 1.042 & 3.306 & 5.487 & & 26.641 \\
    \end{tabular}
    \caption{The regression coefficient, $p$-value and Bayes factor for linear regression of the interannual variability in regime frequency (ERA-Interim onto SEAS5) for all six regimes. In addition, the result of multiple linear regression of the ERA-Interim NAO$-$ frequency against the SEAS5 ensemble mean NAO+ and SB$-$ regime frequencies is shown. Values for both the sequential as well as the ensemble Bayesian approach are shown.}
    \label{tab:regression}
\end{table}

\section{Conclusion and Discussion}
\label{sec:discussion}

A new approach exploiting Bayes Theorem \eqref{eq:bayes} is proposed to obtain a probabilistic regime assignment of the atmospheric state on a given day, based on preexisting definitions of the regimes. The approach combines climatological likelihood functions with prior information from the previous day, using climatological estimates of regime persistence, to obtain a Bayesian regime probability. This sequential probabilistic regime assignment allows for smoother transitions between the regimes and indicates whenever data does not clearly belong to one regime. In contrast to previously studied methods that used a regularised $k$-means clustering algorithm \citep{Falkena2020,Falkena2022} there is no parameter, other than the number of regimes $k$, that has to be selected. Also, the method can be applied in real time as new data comes in. Applying the approach to six wintertime circulation regimes over the Euro-Atlantic sector yields an increase in persistence, without affecting the average regime frequencies for both SEAS5 and ERA-Interim (Figure \ref{fig:compare_sequential}). In addition, for ERA-Interim the 1-day autocorrelation was found to be higher than that obtained using a regularised $k$-means approach containing a persistence constraint \citep{Falkena2020}.
The Bayesian probabilistic regime assignment can help overcome the need for some of the heuristic devices, such as a ``no-regime'' category, that are commonly used in circulation regime studies \citep[e.g.][]{Cassou2005,Grams2017}. The regime probabilities indicate when data cannot be clearly assigned to one regime, whereas the incorporation of prior information ensures persistent regime dynamics.
Here, the focus has been on the regime dynamics within the winter season and on interannual timescales, leaving the challenging problem of seasonality of regimes aside \citep[e.g.][]{breton2022}.

A yet more informative prior for the Bayesian approach can be obtained by continuously updating the prior probabilities by taking information from the full SEAS5 ensemble into account.
Starting from the assumption of approximate stationarity of the ensemble mean regime frequencies at each day, the regime transition matrix is updated.
This update is started from the diagonal of the transition matrix since the persistence dominates the regimes dynamics. The limited availability of data is not sufficient to reliably apply other approaches such as Hidden Markov Models. This updated transition matrix in turn affects the prior probabilities, leading to more pronounced interannual variability for some regimes. 
When considering the interannual variability, the response to three very strong El Ni\~no events in recent decades clearly stands out (Figure \ref{fig:occrate_yearcomp}). During these three winters SB$-$ and NAO$-$ increase in frequency, while NAO+, AR+ and AR$-$ decrease. The signals for AR+, AR$-$ and SB$-$ are enhanced by the ensemble Bayesian approach compared to the sequential method. The signal during La Ni\~na winters is less pronounced, with the increase in NAO+ frequency during 1988-89 standing out most clearly.

This response to ENSO in the SEAS5 ensemble can already be identified using only a 10-member ensemble. The increase in SB$-$ occurrence during El Ni\~no years is a particularly strong signal and is found in nearly all 10-member ensembles considered (Figure \ref{fig:occrate_10members}). Also for NAO+, NAO$-$, AR+ and AR$-$ significant probabilities of detection for peaks or troughs coinciding with El Ni\~no are found. However, here there also is a substantial false alarm ratio indicating that many peaks or troughs in the ensemble occur in non-El Ni\~no years. For La Ni\~na there also is some signal, but not as strong as for El Ni\~no years. These results suggest that one may not need a very large ensemble to identify regime signals in response to ENSO.

We also use a linear regression analysis to identify predictable signals in the observations on interannual timescales. Here, as in \citet{Falkena2022}, NAO+ and SB$-$ were found to be predictable from the SEAS5 ensemble with regression coefficients around one (Table \ref{tab:regression}), suggesting no signal-to-noise deficit for these regimes. The ensemble approach leads to an increase in Bayes factor compared to the sequential method for all regimes, with the largest improvement for NAO+. 

ENSO is certainly part of the reason for the predictable signal found with the regression approach, but it is likely that other processes play a role as well. Previous studies have linked the frequency of Euro-Atlantic circulation regimes to the Madden-Julian Oscillation \citep[e.g.][]{Cassou2008,Straus2015,LeeR2019, Lee2020} and the stratospheric polar vortex \citep[e.g.][]{Charlton-Perez2018,Domeisen2020}, and it would be interesting to see whether the Bayesian approach to regime assignment can aid in better understanding the links between these processes and the regime frequencies. In that respect, the clear improvement in persistence obtained from the sequential method (Figure 5) should be useful for such S2S applications, even if the seasonal averages are not much affected. Information about other climatic processes that are known to affect the regime occurrence can be used to obtain an informative prior for the regime probabilities. For example, knowledge of the states of ENSO or the stratospheric vortex can inform the prior regime probabilities. Such priors can be used for both model ensembles as well as reanalysis datasets and aid in better distinguishing the signal from the noise.

The use of the Bayesian regime assignment approach is not limited to atmospheric circulation regimes, but can be applied to any case in which the data can be separated into two or more regimes. For example, one can think of the two phases of the NAO or the jet latitude \citep{Woollings2010a}. For the application one needs some information on the regime likelihood function and a way to obtain an informative prior. In most cases the latter will be the most challenging and requires a thorough understanding of the processes involved. For circulation regimes a prior based on climatological transition probabilities, which automatically builds in persistence, was shown to be a suitable and natural choice, and incorporating information from a full ensemble enhanced the interannual signal. Depending on the regime process considered other choices for the prior may be more suitable.

\clearpage
%%%%%%%%%%%%%%%%%%%%%%%%%%%%%%%%%%%%%%%%%%%%%%%%%%%%%%%%%%%%%%%%%%%%%
% ACKNOWLEDGMENTS
%%%%%%%%%%%%%%%%%%%%%%%%%%%%%%%%%%%%%%%%%%%%%%%%%%%%%%%%%%%%%%%%%%%%%
\acknowledgments
SKJF was supported by the Centre for Doctoral Training in Mathematics of Planet Earth, with funding from the UK Engineering and Physical Sciences Research Council (EPSRC) (grant EP/L016613/1). The research of JdW has been partially funded by Deutsche Forschungsgemeinschaft (DFG) – SFB1294/1, 318763901. We thank the three reviewers for their constructive feedback.

%%%%%%%%%%%%%%%%%%%%%%%%%%%%%%%%%%%%%%%%%%%%%%%%%%%%%%%%%%%%%%%%%%%%%
% DATA AVAILABILITY STATEMENT
%%%%%%%%%%%%%%%%%%%%%%%%%%%%%%%%%%%%%%%%%%%%%%%%%%%%%%%%%%%%%%%%%%%%%
% 
%
\datastatement
ERA-Interim and SEAS5 hindcast data are publicly available at the ECMWF website.

%%%%%%%%%%%%%%%%%%%%%%%%%%%%%%%%%%%%%%%%%%%%%%%%%%%%%%%%%%%%%%%%%%%%%
% REFERENCES
%%%%%%%%%%%%%%%%%%%%%%%%%%%%%%%%%%%%%%%%%%%%%%%%%%%%%%%%%%%%%%%%%%%%%
%  This shows how to enter the commands for making a bibliography using
%  BibTeX. It uses references.bib and the ametsocV6.bst file for the style.

\bibliographystyle{ametsocV6.bst}
\bibliography{PhD.bib}

\end{document}